\documentclass[fleqn,usenatbib]{mnras}

\usepackage{newtxtext,newtxmath}
\usepackage{natbib}
\defcitealias{thomas2018}{T18}
\defcitealias{thomas2019}{T19}
\usepackage[T1]{fontenc}
\usepackage{ae,aecompl}
\usepackage{pdflscape}
\usepackage{graphicx}	% Including figure files
\usepackage{amsmath}	% Advanced maths commands
\title[The NGC~5466 Stellar Stream]{Uncovering fossils of the distant Milky Way with UNIONS: NGC~5466 and its stellar stream}

\author[J. Jensen et al.]{Jaclyn Jensen$^{1}$\thanks{E-mail: jaclynjensen@uvic.ca},
Guillaume Thomas$^{2,3,4}$, Alan W. McConnachie$^{4,1}$, Else Starkenburg$^{5}$, \newauthor Khyati Malhan$^{6}$, Julio Navarro$^{1}$, Nicolas Martin$^{7}$, Benoit Famaey$^{7}$, Rodrigo Ibata$^{7}$, \newauthor Scott Chapman$^{8}$, Jean-Charles Cuillandre$^{9}$, Stephen Gwyn$^{4}$
\\
$^{1}$Department of Physics \& Astronomy, University of Victoria, Victoria, BC, V8P 1A1, Canada \\
$^{2}$Instituto de Astrof\'isica de Canarias, E-38205 La Laguna, Tenerife, Spain \\
$^{3}$Universidad de La Laguna, Dpto. Astrof\'isica, E-38206 La Laguna, Tenerife, Spain \\
$^{4}$NRC Herzberg Astronomy \& Astrophysics, 5071 West Saanich Road, Victoria, BC, V9E 2E7, Canada \\
$^{5}$Kapteyn Astronomical Institute, University of Groningen, Landleven 12, 9747 AD Groningen, The Netherlands \\
$^{6}$The  Oskar  Klein  Centre  for  Cosmoparticle  Physics,  Department  of Physics,  Stockholm  University,  AlbaNova,  10691  Stockholm,  Sweden \\
$^{7}$l'Universit\'{e} de Strasbourg, CNRS, Observatoire astronomique de Strasbourg, UMR 7550, F-67000 Strasbourg, France \\
$^{8}$Department of Physics and Atmospheric Science, Dalhousie University, 6310 Coburg Rd., Halifax, NS B3H 4R2 Canada \\
$^{9}$AIM, CEA, CNRS, Universit\'{e} Paris-Saclay, Universit\'{e} Paris Diderot, Sorbonne Paris Cit\'{e}, Observatoire de Paris, PSL University, F-91191 Gif-sur-Yvette, \\
France \\
}

\date{Accepted 2021 August 6; Received 2021 July 15; in original form 2021 June 2}

\pubyear{2021}

\begin{document}
\label{firstpage}
\pagerange{\pageref{firstpage}--\pageref{lastpage}}
\maketitle

\begin{abstract}

We examine the spatial clustering of blue horizontal branch (BHB) stars from the $\textit{u}$-band of the Canada-France Imaging Survey (CFIS, a component of the Ultraviolet Near-Infrared Optical Northern Survey, or UNIONS). All major groupings of stars are associated with previously known satellites, and among these is NGC~5466, a distant (16 kpc) globular cluster. NGC~5466 reportedly possesses a long stellar stream, although no individual members of the stream have previously been identified. Using both BHBs and more numerous red giant branch stars cross-matched to $\textit{Gaia}$ Data Release 2, we identify extended tidal tails from NGC~5466 that are both spatially and kinematically coherent. Interestingly, we find that this stream does not follow the same path as the previous detection at large distances from the cluster. We trace the stream across 31$\degr$ of sky and show that it exhibits a very strong distance gradient ranging from 10~$<$~R$_{helio}$~$<$~30~kpc. We compare our observations to simple dynamical models of the stream and find that they are able to broadly reproduce the overall path and kinematics. The fact that NGC~5466 is so distant, traces a wide range of Galactic distances, has an identified progenitor, and appears to have recently had an interaction with the Galaxy's disk, makes it a unique test-case for dynamical modelling of the Milky Way.

% and estimate that it contains at least 8\% of the cluster’s current mass (or 5.5~$\times$~10$^{3}$ M$_{\odot}$ in the stream).

% The NGC~5466 stream also exhibits a very strong distance gradient ranging from 10~$<$~R$_{helio}$~$<$~30~kpc, which we argue is a key . 
%  Contrary to a previous claim, we find that the stream does not follow the same path as the previous detection at large distances from the cluster. 

% We also find that the stream exhibits a very strong distance gradient (ranging from 10~$<$~R$_{helio}$~$<$~30~kpc) such that stars in the leading arm 
% Canada-France Imaging Survey (CFIS) and identify several known satellites within the dataset. 

% This is a simple template for authors to write new MNRAS papers.
% The abstract should briefly describe the aims, methods, and main results of the paper.
% It should be a single paragraph not more than 250 words (200 words for Letters).
% No references should appear in the abstract.
\end{abstract}


%%%%%%%%%%%%%%%%%%%%%%%%%%%%%%%%%%%%%%%%%%%%%%%%%%

%%%%%%%%%%%%%%%%% BODY OF PAPER %%%%%%%%%%%%%%%%%%

\section{Introduction}

In the standard $\Lambda$CDM cosmology, galaxies form hierarchically via a series of mergers (\citealt{white1978,johnston2008}). Larger galaxies accrete smaller stellar systems together with their own globular clusters, and tidal forces act to strip stars from these satellites to form stellar streams. At large distances from the Galaxy's center, dynamical timescales are long such that signatures of these mergers may be observable for many billions of years (\citealt{johnston1996}) and form part of the Galaxy’s “fossil record”. In particular, kinematics and chemical abundances of old stars in these streams provide key insights into the merger history of the Milky Way and the formation of the “proto-Galactic fragments” that have since merged (\citealt{searle1978}). 

% (\citealt{johnston1996,johnston1999,johnston2008,johnston2016})
% Obtaining astrometry and chemical abundances of these halo stars allows us to understand the origins of ancient features 

In recent years, revolutionary large-sky surveys have provided unprecedented perspectives of the Milky Way $-$ for example, the Sloan Digital Sky Survey (SDSS; \citealt{york2000}), Pan-STARRS1 3$\pi$ survey (PS1 3$\pi$; \citealt{chambers2016}), and the Dark Energy Survey (DES; \citealt{des2005}) to name a few. Most notably, the advent of $\textit{Gaia}$ (\citealt{gaia2016}) has been instrumental to measuring the positions and proper motions for billions of stars and developing a detailed map of our Milky Way. $\textit{Gaia}$'s second data release ($\textit{Gaia}$ DR2; \citealt{gaia2018}) ignited a prosperous era for Galactic archaeology, with a host of newly identified substructures. Some key studies using $\textit{Gaia}$ DR2 include (1)~searches for stellar streams (\citealt{malhan2018,mateu2018,ibata2019,necib2020,borsato2020}), (2)~updated globular cluster kinematics (\citealt{baumgardt2019}) (3)~identification of tidal tails from globular clusters (\citealt{bianchini2019,kundu2019,sollima2020,thomas2020}), (4)~new estimates for the Milky Way mass profile (\citealt{cautun2020}), and (5)~unveiling the Galaxy's complex accretion history (\citealt{helmi2018,mackereth2019}), among many other advancements.

Streams around globular clusters have proven to be powerful probes of the Galactic potential (e.g., \citealt{kupper2015,pearson2015,thomas2017,thomas_stream2018,bonaca2018,malhanibata2019}). Due to their lower initial masses and velocity dispersions, globular cluster streams are typically thin, dynamically cold, and extremely sensitive to perturbations from the host potential. Each substructure represents a unique interaction with the Galaxy, yet few are known at large ($>$ 10 kpc) distances. Increasing the number of known systems in this regime, especially those with clear progenitors, can place tighter constraints on the shape and mass of the dark matter halo interior to each distant stream's orbit.

$\textit{Gaia}$’s precise astrometry has  provided many of the pieces necessary to better understand the Galaxy’s interactions with its satellites. However, parallax uncertainties of fainter sources are much less accurate at large distances (\citealt{lindegren2018}); therefore, exploring the distant Milky Way with $\textit{Gaia}$ alone is difficult. In contrast, $\textit{Gaia}$ proper motions remain extremely useful even deep into the stellar halo (\citealt{powell2013}; also see Figure 1 in \citealt{ibata_feh2017}). 

To push $\textit{Gaia}$ out to the distant Galaxy, we have estimated photometric parallaxes of tracer stellar populations using $\textit{u}$-band data from the Canada-France Imaging Survey (CFIS; \citealt{ibata2017}) combined with $\textit{Gaia G}$- (\citealt{gaia2018}) and PS1 3$\pi$ \textit{griz}-bands (\citealt{chambers2016}). At these larger distances, estimated photometric parallaxes based on high-quality ground-based photometry can be very accurate (see the seminal study of \citealt{juric2008}). In \citet{thomas2018,thomas2019} we show that we can obtain distances for stellar populations in the stellar halo ($>$ 10 kpc) that are considerably more accurate than $\textit{Gaia}$ alone. In combination with $\textit{Gaia}$ proper motions, we can therefore explore the dynamical structure of the stellar halo to larger distances than would be otherwise possible.

The paper is organized as follows. In Section \ref{sect:2} we discuss the various large-sky surveys and preliminary data processing of the CFIS tracer populations used in this work. Section \ref{sect:3} details our method to probe the halo via a clustering algorithm that visualizes the hierarchical nature of halo substructures. Analysis of the major features leads us to identify a putative stellar stream around NGC~5466, which forms the focus of the remainder of the paper due to its interesting yet ill-defined properties. We quantify these properties in Section \ref{sect:4}, and in Section \ref{sect:5}, we conduct simple dynamical modelling of this stellar stream. We compare our model to the  observational data and previous work in Section \ref{sect:6}, and summarize our results in Section \ref{sect:7}.

\section{Preliminaries}
\label{sect:2}

\subsection{CFIS, UNIONS, and \textit{Gaia}}

\begin{figure*}
     \centering
    \includegraphics[width=0.75\textwidth]{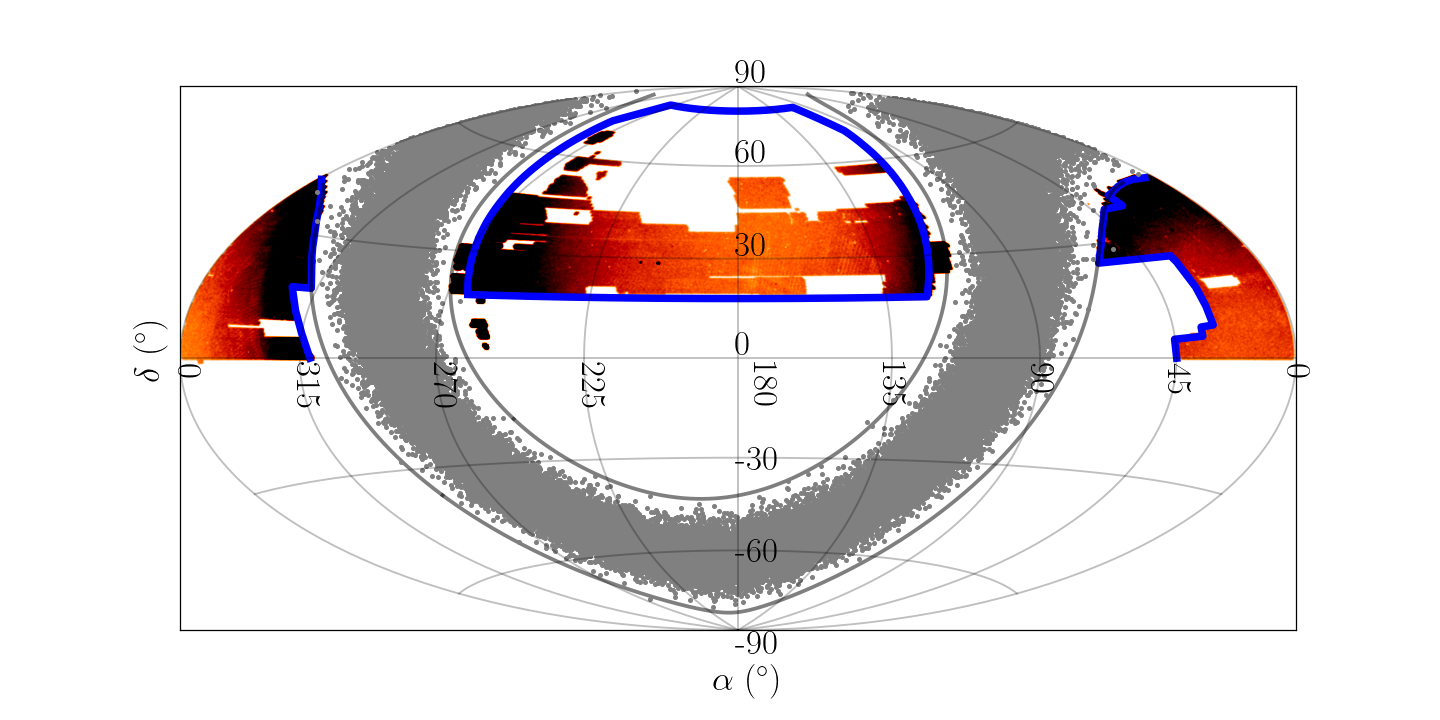}
     \caption{CFIS-$\textit{u}$ equatorial footprint. Blue outlines show the final CFIS area when the survey is complete ($\sim$5,000 deg$^{2}$), while red/orange shaded regions are the fields available at the time of the current study ($\sim$4,000 deg$^{2}$). Grey solid lines are located at $\textit{b}$ = $\pm$20$\degr$ to highlight the boundary of CFIS in Galactic latitude; grey points in this area highlight the approximate position of stars in the disk.}
     \label{fig:CFIS_footprint}
\end{figure*}

    %  Lines show the overall CFIS-$\textit{u}$ footprint ($\sim$10,000 deg$^{2}$), and the blue area shows the footprint at the time of the current study ($\sim$4,000 deg$^{2}$). Also highlighted is the Galactic plane.

%\citealt{prusti2016} same article??
%%%####might cut this out since it was introduced in Section 1
% $\textit{Gaia}$ (\citealt{gaia2016}) provides the astrometry - especially parallaxes and proper motions - that is essential to better understand the dynamical origins of the Galaxy. Gaia astrometry is especially powerful in the “extended Solar neighborhood”, that is for stars which are generally found within 10 kpc or so. Beyond this  distance, $\textit{Gaia}$ parallaxes are, in a relative sense, quite poor (e.g., see see Figure 1 of \citealt{ibata_feh2017}). $\textit{Gaia}$ proper motion estimates in this distance range, however, are still good in a relative sense, but require good distances to convert them to meaningful units. It is with consideration in mind that we started the $\textit{u}$-band imaging program for the Canada-France Imaging Survey (CFIS; \citealt{ibata2017}).

CFIS is an on-going Large Program using the MegaCam imager (\citealt{boulade2003}) at the Canada-France Hawaii Telescope (CFHT). When completed, the survey will have ground-based $\textit{u}$- and $\textit{r}$-band photometry for 10,000 and 5,000 deg$^{2}$ of the northern sky, respectively. The primary motivation for the extensive CFIS-$\textit{u}$ imaging is its power for Galactic studies of nearby stellar populations, in addition to its complementarity to the $\textit{Euclid}$ mission (\citealt{laureijs2011, racca2016}). As demonstrated in \citet{ibata2017} (see their Figure 5), the CFIS $\textit{u}$-band is deeper than SDSS by $\sim$2.7 magnitudes as a result of longer integration times on a larger telescope, which is much more optimized for UV by design (e.g., optical coatings) compared to other facilities. CFIS is focused at Galactic latitudes of $\mid\textit{b}\mid$ $>$ 19$\degr$ and is well-suited for studying the halo. Figure \ref{fig:CFIS_footprint} shows the final overall footprint of the CFIS-$\textit{u}$ component in blue, where the red regions are the area currently available in this work.
% We refer the reader to \cite{ibata2017} for all details relating to survey strategy and data reduction procedures.
 
Recently, the scope of CFIS has expanded alongside multiple other northern hemisphere imaging surveys. Specifically, the Ultraviolet Near-Infrared Optical Northern Survey (UNIONS) is a new consortium of wide-field imaging surveys of the northern hemisphere. UNIONS consists of the CFIS team, members from Pan-STARRS, and the Wide Imaging with Subaru HyperSuprimeCam of the Euclid Sky (WISHES) team. Each group is currently collecting imaging at their respective telescopes: CFHT/CFIS is targeting deep $\textit{u}$- and $\textit{r}$- band photometry, Pan-STARRS is obtaining deep $\textit{i}$- and moderate-deep $\textit{z}$-bands, and Subaru/WISHES is acquiring deep $\textit{z}$. These independent efforts are directed, in part, to securing optical imaging to complement the $\textit{Euclid}$ space mission, although UNIONS is a separate consortium aimed at maximizing the science return of these large and deep ground-based surveys of the northern skies. 

In this contribution, we make use of the UNIONS/CFIS $\textit{u}$-band data only. All CFIS-$\textit{u}$ sources used in this work are cross-matched to the $\textit{griz}$-bands of the PS1 3$\pi$ survey for complete photometric coverage across the optical spectrum (note that PS1 3$\pi$ should not be confused with the new Pan-STARRS $\textit{i}$- and $\textit{z}$-band imaging being obtained as part of the UNIONS effort). Astrometry for these sources are obtained from the second $\textit{Gaia}$ data release (\citealt{gaia2018}), as this work preceded the arrival of $\textit{Gaia}$ eDR3 (\citealt{gaia2020}).

\subsection{Tracer stellar populations}
\label{sect:pops}

The $\textit{u}$-band is exceptionally useful for the study of nearby stellar populations. For example, a star’s absolute magnitude is sensitive to its metallicity, and many metal lines are found in the UV-blue region of the spectrum. This fact is particularly useful to photometrically identify target populations and derive basic parameters. In this work, we target specific tracer stellar populations for which the absolute magnitudes are reasonably well-constrained. The resulting distances, when paired with the excellent proper motions from $\textit{Gaia}$ DR2, gives us a more complete kinematic view of the outer Galaxy than is possible when using solely $\textit{Gaia}$.

% \citet[hereafter, \citetalias{thomas2018}]{thomas2018}

The first stellar population used in this study are the blue horizontal branch stars (BHBs) that were identified in \citet[hereafter \citetalias{thomas2018}]{thomas2018}. BHBs are an ideal tracer to probe the stellar halo for substructure: these bright A-type giants have relatively stable absolute magnitudes ($M_g$ $\sim$ 0.5 $-$ 0.7 mag; \citealt{deason2011}), allowing us to trace them out to large distances. Using CFIS-$\textit{u}$ and PS1-$\textit{griz}$ extinction-corrected bands, \citetalias{thomas2018} identified A-type stars using a sequence of color-color cuts. BHB stars were then disentangled from contaminating blue stragglers via a random forest classifier, producing a sample of $\sim$10,200 BHBs with $\sim$25\% contamination from blue straggler stars. The absolute magnitudes were derived using the calibration from \citet{deason2011}, where M$_{g}$ is a function of ($\textit{g}_0$ $-$ $\textit{r}_0$). In \citetalias{thomas2018}, the heliocentric (photometric) distances are estimated from M$_{g}$ which are shown to be accurate to $\sim$10\%, extending out to $\sim$220 kpc.

The second tracer population used in this work is from the catalog of “dwarfs” and “giants” in \citet[hereafter \citetalias{thomas2019}]{thomas2019}. Briefly, \citetalias{thomas2019} implemented a machine learning scheme to classify stars as either main sequence (MS/“dwarfs”) or red giant branch (RGB/“giants”), using SEGUE spectra and $\textit{Gaia}$ photometry and parallaxes as a training set. For both the dwarfs and the giants, photometric metallicities ([Fe/H]) and absolute magnitudes in $\textit{Gaia G}$-band (M$_{\textit{G}}$) were estimated. The initial classification assigns a probability to each star that it is either a dwarf or giant (such that P$_{dwarf}$~+~P$_{giant}$~=~1) based on its color using the combined photometry of CFIS-PS13$\pi$-$\textit{Gaia G}$. The algorithm successfully identifies 70\% of metal-poor giants with [Fe/H] $<$ $-$1.2 dex. Then, each population is run through their own set of Artificial Neural Networks, which serve to estimate [Fe/H] and M$_{\textit{G}}$ from the training set. \citetalias{thomas2019} show that the uncertainties on the photometric metallicities and distances for the giants are approximately 0.3 dex and $<$25\%, respectively. These authors also note that more metal-rich giants are often misidentified, resulting in a significant drop in completeness for [Fe/H] $>$ $-$1 dex. This minimally affects our current study, as we are primarily concerned with giants in the metal-poor regime.

In what follows, we start with a dataset composed of all likely giants (those with P$_{giant}$~$>$~50\%; of order $\sim$600,000 sources). We then remove any potential background galaxies using the PS1 criterion in \citet{farrow2014}, $\mid r_{PSF} - r_{ap}\mid$ $<$ 0.05. While this method becomes unreliable for stars fainter than $\textit{r}_{PSF}$ > 21 mag, 99.9\% of stars in this dataset are brighter than this limit $-$ given that we are limited by the depth of $\textit{Gaia}$ at $\textit{G}$ $\simeq$ 21 mag. As discussed, an artefact of the method of \citetalias{thomas2019} is the misidentification of more metal-rich stars. Following their recommendations, we remove this contamination by restricting the uncertainties of the predicted absolute magnitudes, $\delta$M$_{G,pred}$ $\leq$ 0.5 mag (where $\delta$M$_{G,pred}$ includes the photometric and systematic errors added in quadrature). This reduces our sample size to $\sim$201,000 RGBs. Finally, we apply a restriction to the $\textit{Gaia}$ parallaxes, in order to avoid nearby stars in the Solar neighborhood (\citealt{lindegren2018}). This equates to a parallax cut at $>$ 0.2 mas, or equivalently, removing stars whose heliocentric distances are less than 5 kpc. \citet{gaia2018} also recommend a zero-point correction of $-$0.03 mas to account for the global parallax offset. Therefore, we adopt:

\begin{equation}
    \frac{1}{\pi + 0.03 \; \text{mas}} > 5 \; \text{kpc}
\end{equation}

\noindent This provides us with a total sample of $\sim$103,000 RGB stars with positions, proper motions, distances, and metallicities.

\section{Searching the halo for stellar substructures}
\label{sect:3}

In this section, we begin by examining the clustering of the outer halo, taking advantage of the relatively precise distances of the BHB sample. We identify several major substructures, all of which are well-known, including the NGC~5466 globular cluster. We identify a group of co-moving stars surrounding NGC~5466 and attempt to better trace its extension using improved statistics (but less precise distances) of the \citetalias{thomas2019} RGBs.

% After identifying a group of co-moving stars around the globular cluster NGC~5466, we attempt to better trace this extended feature using improved statistics (but less precise distances) of the \citetalias{thomas2019} RGBs.

\subsection{Examining the clustering of blue horizontal branch stars}

\subsubsection{The OPTICS spatial clustering algorithm}

\begin{figure*}
    \centering
    \includegraphics[width=0.31\textwidth]{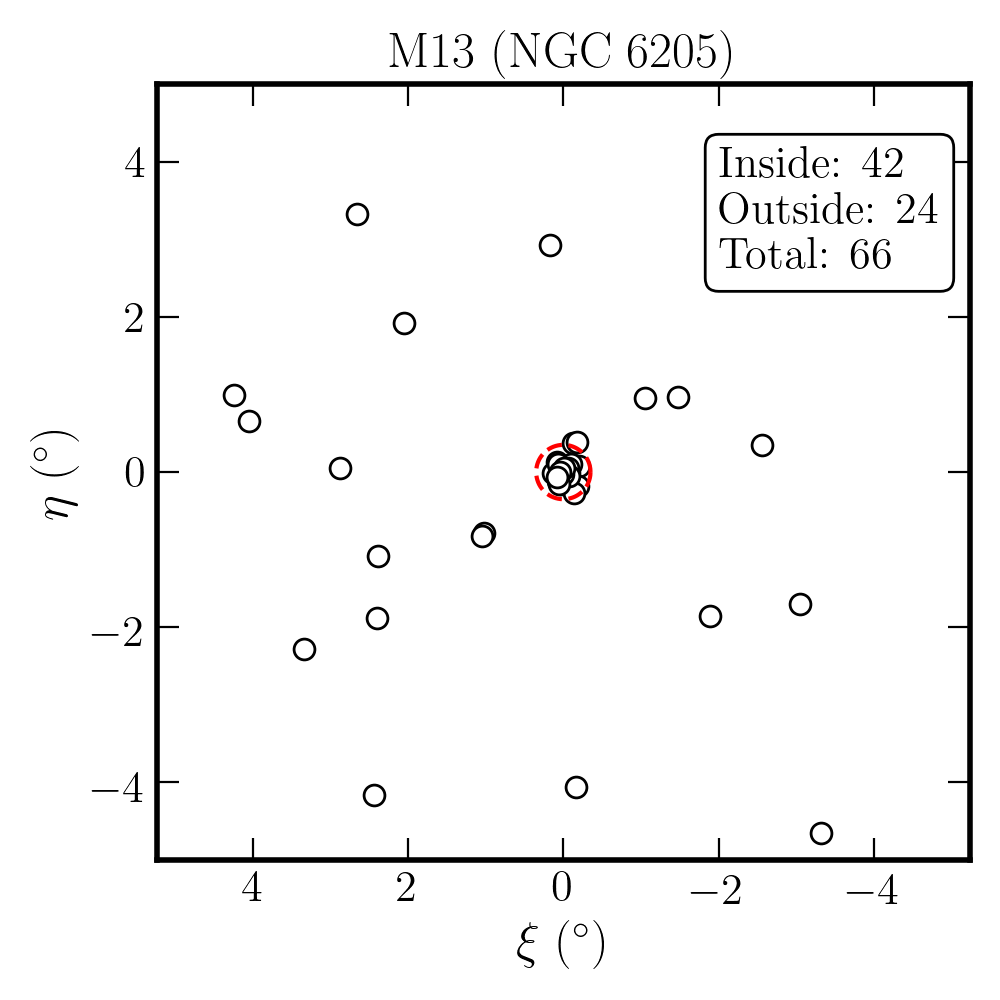}
    \includegraphics[width=0.31\textwidth]{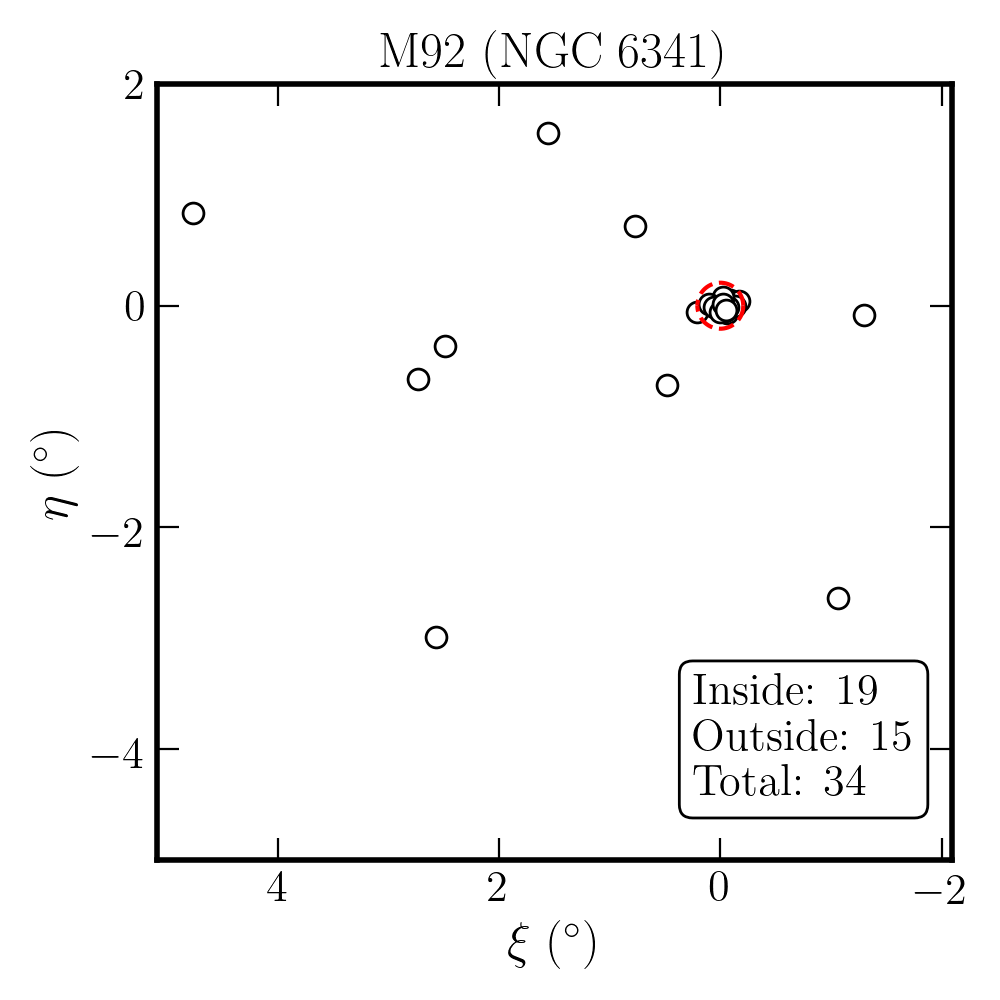}
    \includegraphics[width=0.31\textwidth]{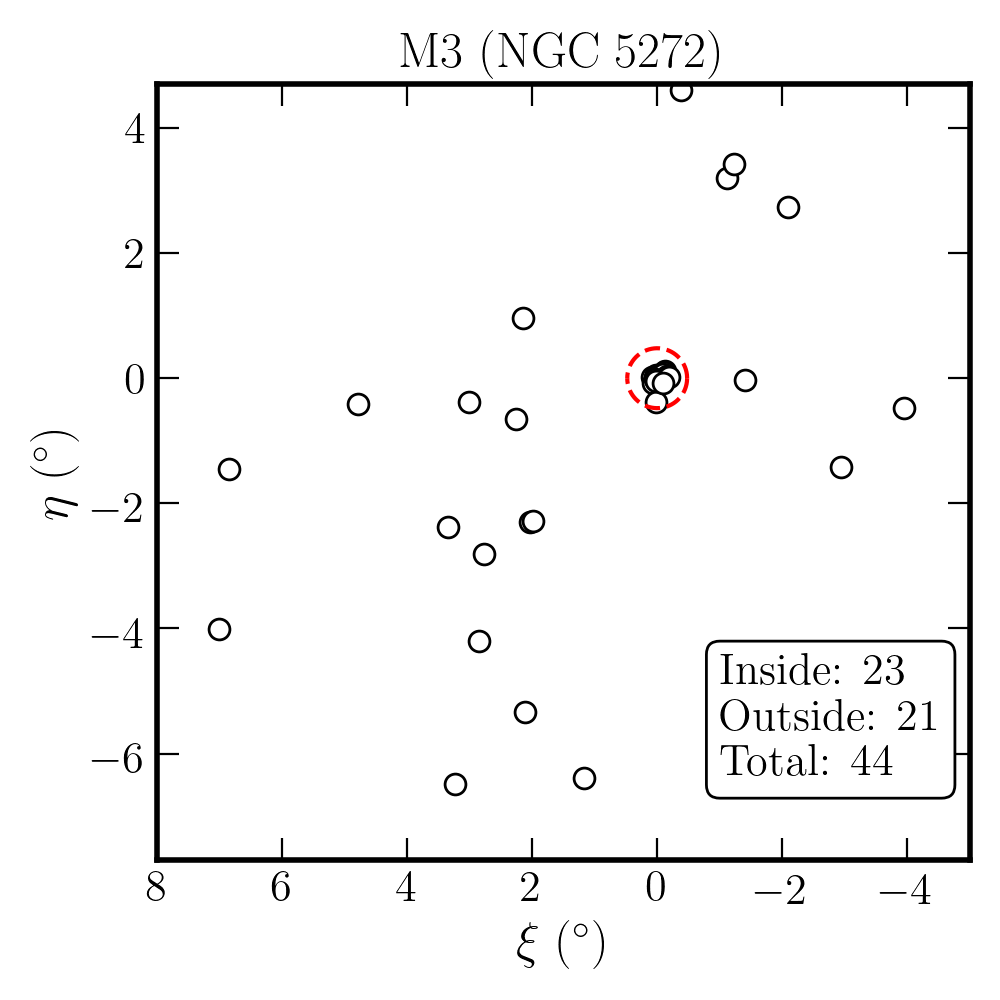}
    
    \includegraphics[width=1\textwidth]{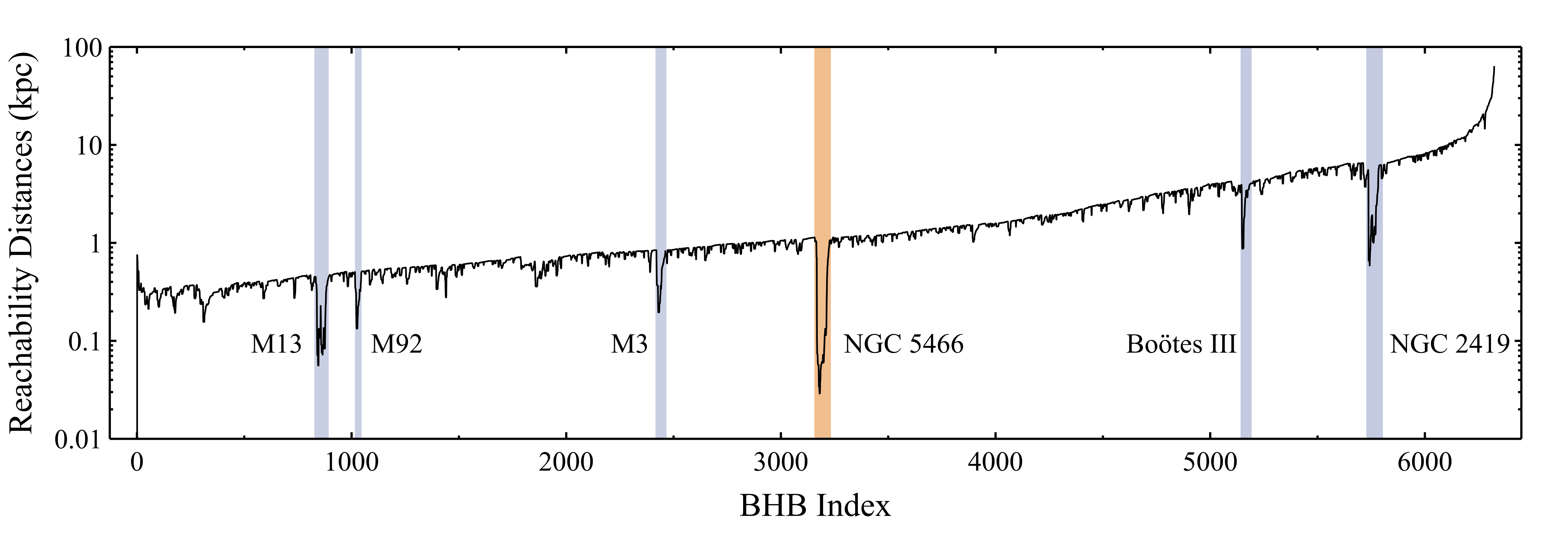}
    
    \includegraphics[width=0.31\textwidth]{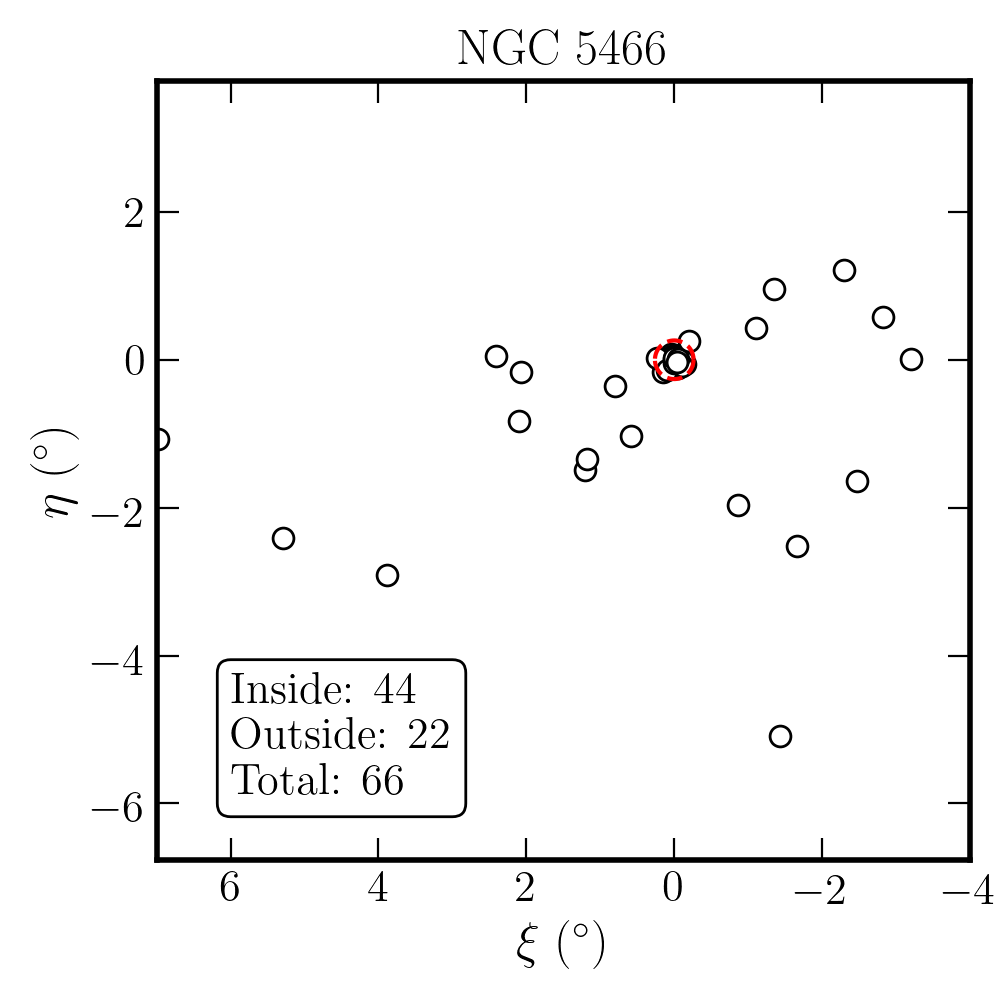}
    \includegraphics[width=0.31\textwidth]{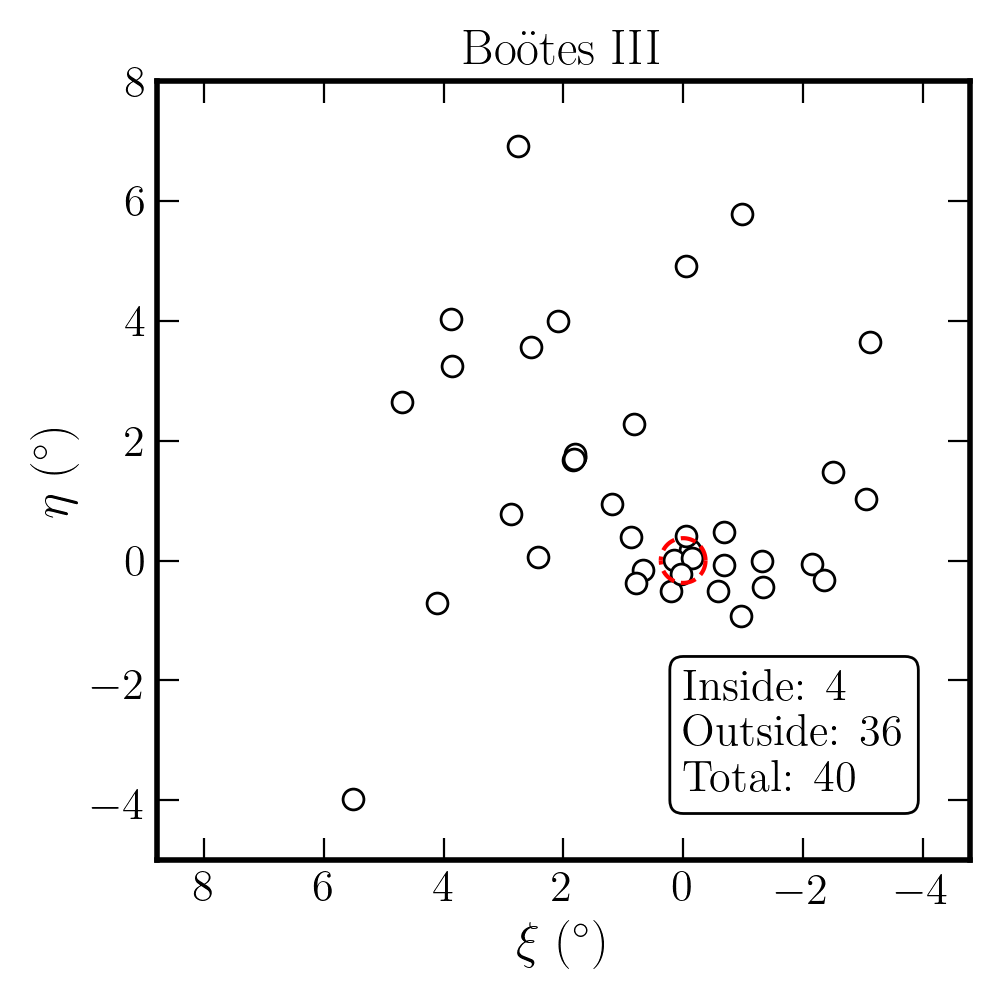}
    \includegraphics[width=0.31\textwidth]{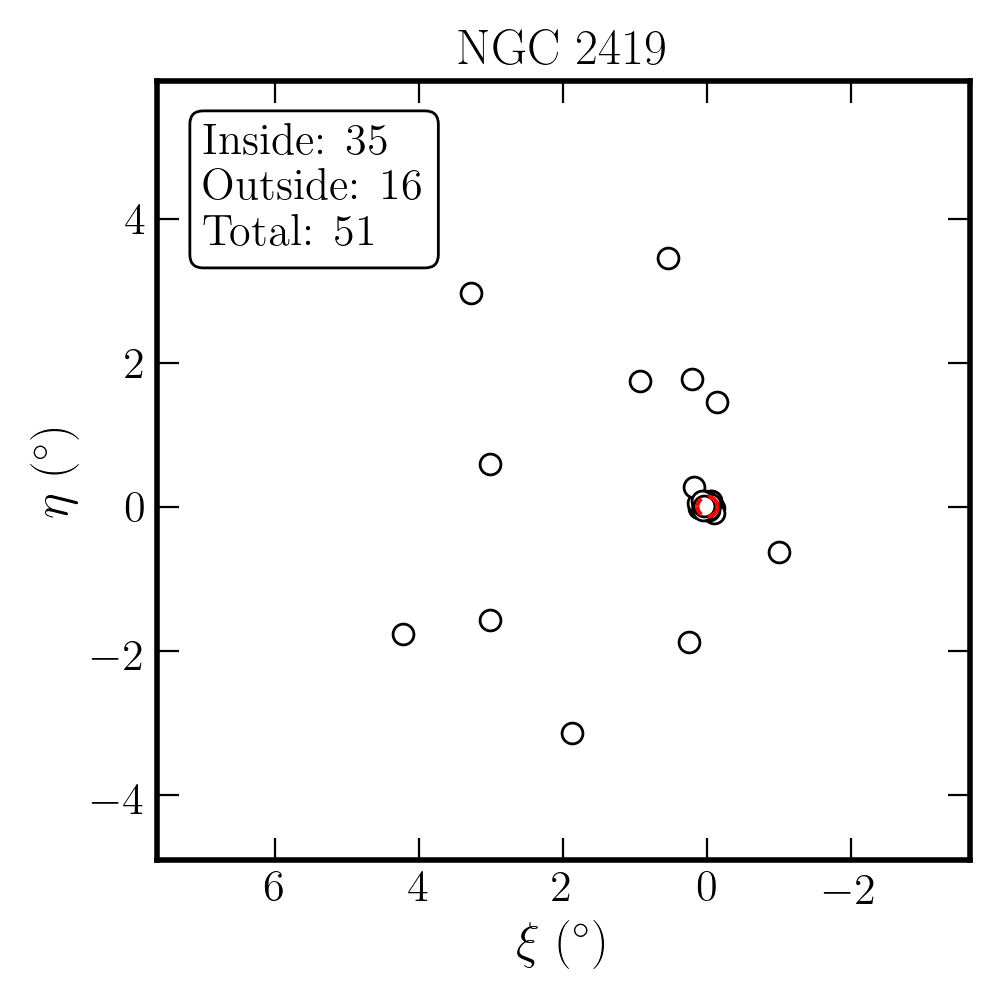}

    \caption{The dendrogram produced by OPTICS (central panel) with major substructures highlighted. Top and bottom rows show the tangent plane projection of the stars associated with each of these substructures, which correspond to known globular clusters or dwarf galaxies. For each, the red dashed circle corresponds to the literature measurement of the cluster's tidal radius (or in the case of Bo\"otes III, its half-light radius). Statistics for each object are listed describing the number of BHBs inside and outside r$_{t}$ (or r$_{h}$).}
    \label{fig:OPTICS}
\end{figure*}

The identification of halo substructures within the spatial distribution of stars requires an algorithm that can identify the clustering of points without strong restrictions on the range of allowable sizes and shapes, and which ideally allows for hierarchical distributions (i.e., substructures within substructures). Here, we opt to use the density-based algorithm known as OPTICS (\textbf{O}rdering \textbf{P}oints \textbf{t}o \textbf{I}dentify \textbf{C}lustering \textbf{S}tructure; \citealt{ankerst1999}).

OPTICS is similar in methodology to other connectivity-based clustering algorithms such as DBSCAN (\citealt{ester1996}). However, OPTICS is more optimal for our study as it does not (a) automatically segregate the data into clusters, (b) requires only one user-specified parameter (N$_{min}$, the minimum number of points that can define a substructure) further limiting potential biases imparted by the algorithm, and (c) produces a useful dendrogram for visualizing the clustering within a dataset (known as the “Reachability Diagram”). OPTICS works by reordering data such that points in the same neighborhood are physically close together in the dendrogram. Overdensities in this plot stand out as “valleys”, or neighborhoods in which the density is clearly higher than its surroundings.

The application of OPTICS to astronomical datasets is, to the best of our knowledge, relatively new. It has been tested with simulations of the Milky Way stellar halo (\citealt{sansfuentes2017}), and has been applied to quantify the properties of the halo of M31 (\citealt{mcconachie2018}). Most recently, \citet{oliver2020} examine in detail how to best apply OPTICS to the study of stellar halos. We refer to these papers for complete details on OPTICS.

\subsubsection{Identifying the most prominent substructures}
\label{sect:312}

First, we transform all BHB positions to the Galactocentric frame using their photometric distances and equatorial positions. For this calculation, we assume the Sun's position is (X, Y, Z)$_{\odot}$ = ($-$8.122,~0,~0.025)~kpc (\citealt{gravity2019,juric2008}). Then, for stars in the northern Galactic region ($\textit{b}$~$>$~20$\degr$; between R.A.~=~[270$\degr$, 90$\degr$] of Figure \ref{fig:CFIS_footprint}), we apply OPTICS to their (X, Y, Z) Galactocentric positions, assuming the minimum number of points to classify a substructure is $N_{min}$ $\geq$ 6. We choose this value for $N_{min}$ as it is sufficient to identify known structures such as globular clusters without producing much noise in the Reachability Diagram. 

% \begin{figure*}
%     \centering
%     \includegraphics[width=0.31\textwidth]{Figures/M13OPTICS.png}
%     \includegraphics[width=0.31\textwidth]{Figures/M92OPTICS.png}
%     \includegraphics[width=0.31\textwidth]{Figures/M3OPTICS.png}
    
%     \includegraphics[width=1\textwidth]{Figures/NGC_RD_Order.pdf}
    
%     \includegraphics[width=0.31\textwidth]{Figures/NGC5466OPTICS.png}
%     \includegraphics[width=0.31\textwidth]{Figures/BooiiiOPTICS.png}
%     \includegraphics[width=0.31\textwidth]{Figures/NGC2419OPTICS.png}

%     \caption{The dendrogram produced by OPTICS (central panel) with major substructures highlighted. Top and bottom rows show the tangent plane projection of the stars associated with each of these substructures, which correspond to known globular clusters or dwarf galaxies. For each, the red dashed circle corresponds to the literature measurement of the cluster's tidal radius (or in the case of Bo\"otes III, its half-light radius). Statistics for each object are listed describing the number of BHBs inside and outside r$_{t}$ (or r$_{h}$).}
%     \label{fig:OPTICS}
% \end{figure*}

The resulting Reachability Diagram is shown in the central panel of Figure \ref{fig:OPTICS}. The x-axis represents the order, or index, of the BHBs within the reorganized dataset. As a result, points physically located near other points in a neighborhood appear close together on the x-axis.

The y-axis of Figure \ref{fig:OPTICS} shows the reachability-distances (RDs) for each star, or the physical distance of a BHB to its associated cluster. RDs essentially give an estimate for the spatial scale of each neighborhood, where clusters in the dataset are represented as valleys. These structures show stars that have small RDs compared to then background, and therefore can be considered a substructure relative to their surroundings.

OPTICS does not automatically define clusters, so we use a custom algorithm also used in \citet{mcconachie2018} to identify prominent valleys. The most significant structures identified with this algorithm are highlighted in Figure \ref{fig:OPTICS}, also corresponding to the largest features identifiable by eye. Of these six highlighted valleys, five are associated to known globular clusters (\citealt[][2010 edition]{harris1996}) and one to a tidally disrupting dwarf galaxy (Bo{\"o}tes III; \citealt{grillmair2016}). The surrounding panels of Figure \ref{fig:OPTICS} show tangent-plane projections of the member stars in each feature, centered on each satellite’s position. We show the King tidal radii (r$_{t}$; \citealt{moreno2014}) in each panel as a red dashed circle; Bo{\"o}tes III is the only exception where we instead show the half-light radius measured by \citet{carlin2009}. In each case, a significant number of BHBs identified as being associated to the main structure lie well beyond the tidal radii of the satellite. If any of these OPTICS-identified stars are indeed actually associated with the satellite, this would suggest extended features around each object. 

% In addition to these highlighted valleys are spurious detections, and smaller features visible in the Reachability Diagram. The broadest of these sub-features is located between the M~92 and M~3 labels on the diagram (index values ranging from approximately 1800 $-$ 2000, corresponding to nearly 200 BHBs). We examined this structure and found it to be an artefact from the cone of observation caused by the CFIS limit in Galactic latitude above the disk. 

Below, we briefly summarize the literature associated with each satellite including any previous detections of tidal debris:

\begin{itemize}

    \item M~13 (NGC~6205) was found to have a “halo of unbound stars” by \citet{lehmann1997} who first examined its King profile (\citealt{king1962}) and identified a surface density excess at the outer regions of the cluster. Later, \citet{leon2000} similarly found an extension of stars towards the Galactic center; however, these stars all lie within the cluster’s estimated tidal radius. 

    \item Tidal tails from the M~92 cluster (NGC~6341) have recently been discovered in two separate detections by \citet{sollima2020} and \citet{thomas2020}, with the latter paper identifying lengthy tails extending over $\sim$17$\degr$.

    \item Two papers (\citealt{leon2000,grillmair2006}) searched for stripping surrounding the M~3 globular cluster (NGC~5272) but do not find any evidence of disruption.

    \item NGC~2419 is a cluster which likely originated from the Sagittarius dwarf galaxy (\citealt{bellazzini2020}). At a distance of $\sim$83 kpc (\citealt[][2010 edition]{harris1996}), the current tidal forces experienced by this cluster will be quite weak, and no tails have previously been reported.

    \item The Bo{\"o}tes III dwarf galaxy is currently being tidally disrupted, and is likely the progenitor of the Styx stellar stream (\citealt{grillmair2009,carlin2009,carlin2018}).

    \item Evidence for mass loss from the NGC~5466 globular cluster was first presented in \citet{pryor1991} and \citet{lehmann1997}. Two detections of tidal tails from this cluster were observed in SDSS, but differed in length: \cite{belokurov2006} identified extra-tidal stars out to 4$\degr$, while \citet{grillmair2006} found very extended tails stretching over $\sim$45$\degr$ of sky using a matched filter method. \citet{fellhauer2007} and \citet{lux2012} modeled the disruption of the cluster based upon the \citet{grillmair2006} detection. These dynamical studies were only able to reproduce the path of the tails over a segment of the matched filter map up to R.A.~$\lesssim$~192$\degr$.

\end{itemize}

The search for new Galactic structures in $\textit{Gaia}$ DR2 has been plentiful, and many groups have mined this catalogue for stellar streams; for example, \citet{malhan2018} and \citet{ibata2019} found 13 new streams using the $\textsc{streamfinder}$ algorithm applied to $\textit{Gaia}$ 5-D kinematics, \citet{mateu2018} identified 14 candidate streams searching over great circles of RR Lyrae, and additional works identifying substructures were conducted by \citet{helmi2017}, \citet{necib2019}, \citet{borsato2020} (to name a few others). During the final stages of preparation of this manuscript, \cite{ibata2020} applied $\textsc{streamfinder}$ to $\textit{Gaia}$ eDR3 and identified some stars associated with a putative stream from NGC~5466 extending $\sim$18$\degr$ on the sky, although no follow-up or commentary was provided.

\subsubsection{A closer look at NGC~5466}
\label{sect:tidal_debris}

\begin{table}
    \centering
    \renewcommand{\arraystretch}{1.3}
    \caption{Observational properties of NGC~5466 summarized from the literature. (1) \citet[][2010 edition]{harris1996}, (2) \citet{moreno2014}, (3) \citet{pryor1991,fellhauer2007}, (4) \citet{baumgardt2019}.}
    \begin{tabular}{c|c|c}
        \hline
         Parameter & Value & Source \\
         \hline\hline
         R.A ($\alpha$) & 211.3637$^\circ$ & (1) \\
         Dec. ($\delta$) & 28.5344$^\circ$ & \\
         Distance (R$_{helio}$) & 16.0 $\pm$ 0.4 kpc & \\
         Concentration (c) & 1.04 $\pm$ 0.2 & \\
         Half-light radius (r$_{h}$) & 2.3 $\pm$ 0.07$\arcmin$ & \\
         Core radius (r$_{c}$) & 1.43 $\pm$ 0.1$\arcmin$ & \\
         Metallicity ($[$Fe/H$]$) & $-$1.98 $\pm$ 0.09 dex & \\
        \hline         
        King tidal radius (r$_{t}$) & 72.98 pc & (2) \\
        \hline
        Stellar mass & $\sim$5 $\times$ $10^{4}$ M$_\odot$ & (3) \\
        % Mass & 5$\textsc{e}$4 M$_{\odot}$ & (3) \\
        \hline
         Proper motion in R.A. ($\mu_{\alpha}$*) & $-$5.41 $\pm$ 0.01 mas\,yr$^{-1}$ & (4)  \\
         Proper motion in Dec. ($\mu_{\delta}$) & $-$0.79 $\pm$ 0.01 mas\,yr$^{-1}$ & \\
         Radial Velocity (v$_{r}$) & 106.93 $\pm$ 0.18 km\,s$^{-1}$ & \\
         \hline
        %  Pericenter (r$_{peri}$) & 6.4 kpc & This work \\
        %  Apocenter (r$_{apo}$) & 43.0 kpc & \\
        %  Eccentricity ($\epsilon$) & 0.74 & \\
        %  L$_{z}$ & & \\
        %  \hline
    \end{tabular}
    \label{tab:ngc5466}
\end{table}

\begin{figure*}
    \centering
    \includegraphics[width=0.35\textwidth]{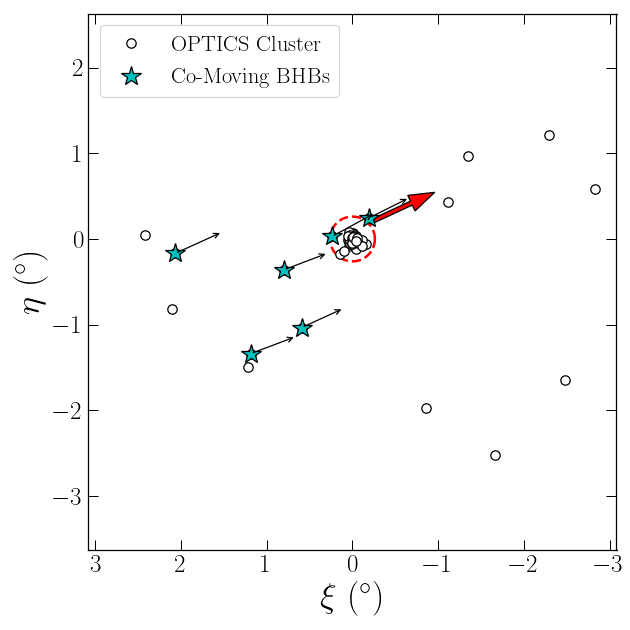}
    \includegraphics[width=0.35\textwidth]{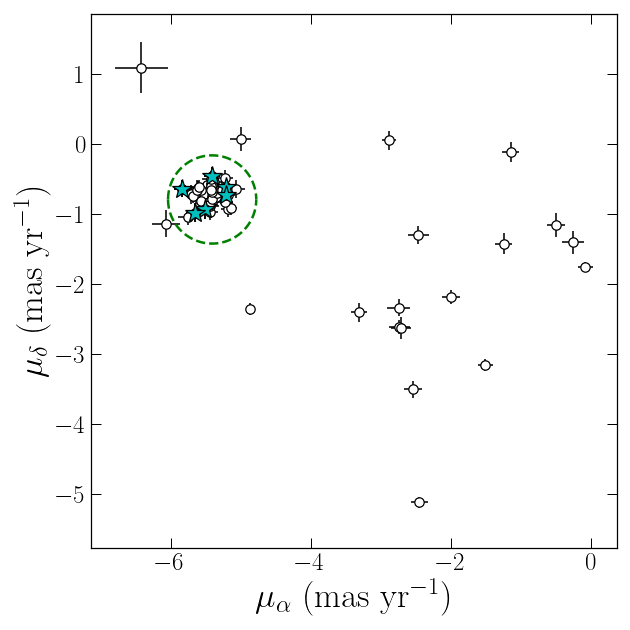}
    \caption{Left: Tangent plane projection of the OPTICS BHBs identified as part of the NGC~5466 group (orange valley in Figure \ref{fig:OPTICS}). The red dashed circle corresponds to the tidal radius of the cluster, and the red arrow represents the proper motion of the cluster (\citealt{baumgardt2019}) corrected for Solar reflex motion. Right: Proper motion vector point diagram for the OPTICS BHBs of NGC~5466. The cluster of stars within the green circle correspond to stars moving with the same systemic proper motion as the cluster. We highlight the BHBs corresponding to stars moving with the cluster (outside r$_t$) as cyan stars in both panels. Their (Solar reflex-corrected) proper motion vectors are also shown as black arrows in the tangent plane.}
    \label{fig:NGC5466_BHBs}
\end{figure*}

% NGC~5466 showed the most compelling evidence for extra tidal stars. 

% \begin{table}
%     \centering
%     \renewcommand{\arraystretch}{1.3}
%     \caption{Observational properties of NGC~5466 summarized from the literature. (1) \citet[][2010 edition]{harris1996}, (2) \citet{moreno2014}, (3) \citet{pryor1991,fellhauer2007}, (4) \citet{baumgardt2019}.}
%     \begin{tabular}{c|c|c}
%         \hline
%          Parameter & Value & Source \\
%          \hline\hline
%          R.A ($\alpha$) & 211.3637$^\circ$ & (1) \\
%          Dec. ($\delta$) & 28.5344$^\circ$ & \\
%          Distance (R$_{helio}$) & 16.0 $\pm$ 0.4 kpc & \\
%          Concentration (c) & 1.04 & \\
%          Half-light radius (r$_{h}$) & 2.3$\arcmin$ & \\
%          Core radius (r$_{c}$) & 1.43\arcmin & \\
%          Metallicity ($[$Fe/H$]$) & $-$1.98 dex & \\
%         \hline         
%         King tidal radius (r$_{t}$) & 72.98 pc & (2) \\
%         \hline
%         Stellar mass & 5 $\times$ $10^{4}$ M$_\odot$ & (3) \\
%         % Mass & 5$\textsc{e}$4 M$_{\odot}$ & (3) \\
%         \hline
%          Proper motion in R.A. ($\mu_{\alpha}$*) & -5.41 $\pm$ 0.01 mas\,yr$^{-1}$ & (4)  \\
%          Proper motion in Dec. ($\mu_{\delta}$) & -0.79 $\pm$ 0.01 mas\,yr$^{-1}$ & \\
%          Radial Velocity (v$_{r}$) & 106.93 $\pm$ 0.18 km\,s$^{-1}$ & \\
%          \hline
%         %  Pericenter (r$_{peri}$) & 6.4 kpc & This work \\
%         %  Apocenter (r$_{apo}$) & 43.0 kpc & \\
%         %  Eccentricity ($\epsilon$) & 0.74 & \\
%         %  L$_{z}$ & & \\
%         %  \hline
%     \end{tabular}
%     \label{tab:ngc5466}
% \end{table}

NGC~5466 is a relatively distant cluster (R$_{helio}$ = 16 kpc). Therefore, a large stream from this satellite would constitute a valuable dynamical tracer of the gravitational potential for the distant Galaxy. A review of the literature to date on this satellite highlights some discrepant claims surrounding the putative stream's properties, which has also proven difficult to model satisfactorily. The extensive coverage of CFIS therefore seems well-suited to better determine the stream's properties. It is with these considerations in mind that we decided to conduct a more complete examination of this structure. We summarize the relevant observational parameters of NGC~5466 in Table~\ref{tab:ngc5466}.

Figure \ref{fig:NGC5466_BHBs} shows the resulting OPTICS grouping for NGC~5466. The left panel shows a zoom-in of the tangent plane as seen in Figure \ref{fig:OPTICS} and the right shows their associated proper motions and uncertainties as reported by $\textit{Gaia}$ DR2. Proper motion errors are corrected by a factor of 1.1, as these values are typically underestimated by 7~$-$~10\% for fainter sources ($\textit{G}$ $>$ 16 mag; \citealt{lindegren2018}). The centroid of the green circle is located at the cluster’s proper motion ($\mu_{\alpha}*$, $\mu_{\delta}$) = ($-$5.41, $-$0.79) mas\,yr$^{-1}$ as derived by \citet{baumgardt2019}. We also represent the cluster's motion in the tangent plane as the red vector, after correcting for Solar reflex motion assuming a distance to the cluster of 16.0 kpc (\citealt[][2010 edition]{harris1996}). For this calculation, we adopt Local Standard of Rest (LSR) values from \citet{schonrich2010} ([U,~V,~W]$_{\odot}$ = [11.1,~12.24,~7.25] km\,s$^{-1}$) assuming the circular velocity at the Sun is 229 km\,s$^{-1}$ (\citealt{eilers2019}) and the Sun's position is the same as in Section \ref{sect:312}. 

% For this calculation, we adopt values from \citet{schonrich2010} for the Local Standard of Rest (LSR): [U,V,W]$_{\odot}$ = [11.1, 12.24, 7.25] km\,s$^{-1}$, \textcolor{red}{}

% For this calculation, we adopt values from \citet{schonrich2010} for the Local Standard of Rest (LSR): [U,V,W]$_{\odot}$ = [11.1, 12.24, 7.25] km\,s$^{-1}$, assuming the Sun's position in the Galactocentric frame is (X,~Y,~Z) = ($-$8.122, 0, 0.025) kpc (\citealt{gravity2019,juric2008}), and the circular velocity at the Sun is 229 km\,s$^{-1}$ (\citealt{eilers2019}).  

The right panel of Figure \ref{fig:NGC5466_BHBs} shows a clear clustering of points corresponding to the systemic proper motions of NGC~5466. Interestingly, there are six BHBs outside r$_{t}$ (red dashed circle) whose proper motions are consistent with stars in the cluster’s main body, shown in both panels as the cyan stars. As done previously with the cluster's proper motion vector, we correct these cyan stars for Solar reflex motion using their photometric distances and previously assumed values for the motion and position of the Sun. We overlay their corrected proper motion vectors on the tangent plane to show these BHBs are clearly moving in a similar fashion as the globular cluster itself, suggesting they are an extra-tidal population from NGC~5466. Our detection of a co-moving group of BHBs is consistent with earlier findings of the stream from \citet{belokurov2006} and \citet{grillmair2006}. However, the BHB population by itself is too sparse to clearly define the path of this putative stream.

% vectors are shown for these candidates in the tangent plane. These BHBs are clearly moving in a similar fashion as the globular cluster itself, suggesting they are an extra-tidal population from NGC~5466. 

% \begin{figure*}
%     \centering
%     \includegraphics[width=0.35\textwidth]{Figures/ngc5466_reversedxi.png}
%     \includegraphics[width=0.35\textwidth]{Figures/ngc5466_pm.png}
%     \caption{Left: Tangent plane projection of the OPTICS BHBs identified as part of the NGC~5466 group (orange valley in Figure \ref{fig:OPTICS}). The red dashed circle corresponds to the tidal radius of the cluster, and the red arrow represents the proper motion of the cluster (\citealt{baumgardt2019}) corrected for Solar reflex motion. Right: Proper motion vector point diagram for the OPTICS BHBs of NGC~5466. The cluster of stars within the green circle correspond to stars moving with the same systemic proper motion as the cluster. We highlight the BHBs corresponding to stars moving with the cluster (outside r$_t$) as cyan stars in both panels. Their (Solar reflex-corrected) proper motion vectors are also shown as black arrows in the left panel.}
%     \label{fig:NGC5466_BHBs}
% \end{figure*}

\subsection{Using red giant branch stars to trace NGC~5466}

\subsubsection{Filtering the red giant branch stars}

\begin{figure}
    \centering
    \includegraphics[width=0.45\textwidth]{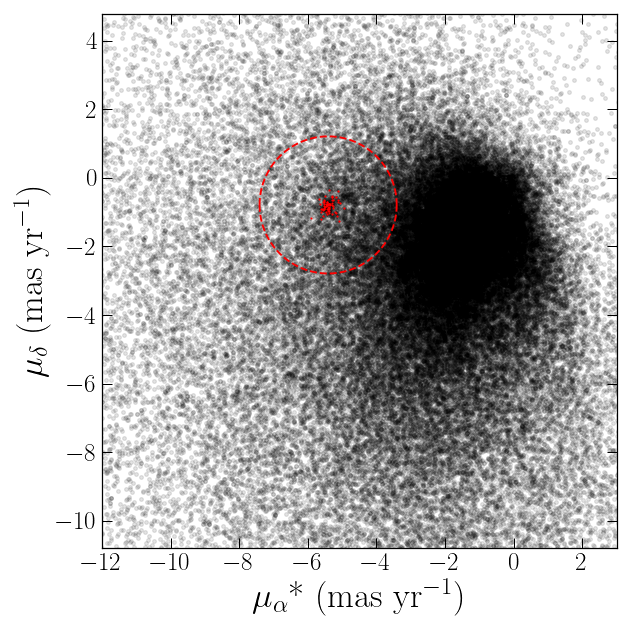}
    \caption{Vector point diagram of all giants (P$_{giant}$ $>$ 50\%) in the CFIS dataset. Stars within 1~$-$~2r$_h$ of the cluster are shown as red points. The red dashed circle is the 2~mas~yr$^{-1}$ boundary chosen to select stars with broadly similar proper motions to the globular cluster.}
    \label{fig:pm_cut}
\end{figure}

\begin{figure}
    \centering
    \includegraphics[width=0.47\textwidth]{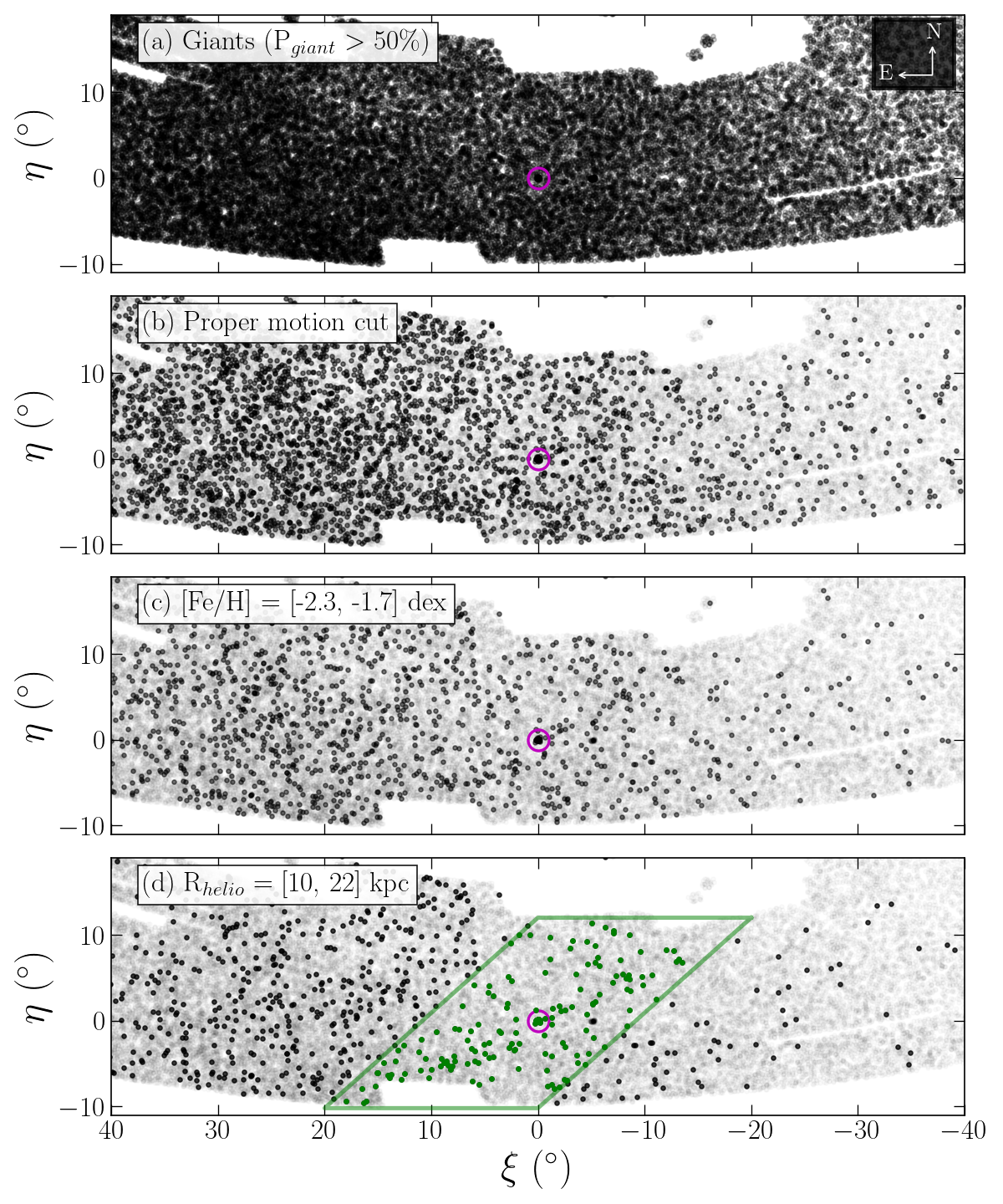}
    \caption{Tangent plane projection of giants in CFIS, centered on NGC~5466 (magenta circle). \textbf{(a):} Giants with P$_{giant}$ $>$ 50\%. \textbf{(b):} Grey points are the full sample as shown in (a), black points are the sample after filtering for proper motion. \textbf{(c):} Stars remaining after filtering for metallicity,  [Fe/H]~=~[$-$2.3,~$-$1.7]~dex. \textbf{(d):} Stars remaining after filtering for distance, R$_{helio}$~=~[10,~22]~kpc. Data within the green area are retained for further analysis.}
    \label{fig:Giants_Cuts}
\end{figure}

Here, we use the more numerous RGB stars to better trace the extended NGC~5466 system. We begin with the sample selected in Section \ref{sect:pops} of $\sim$103,000 stars, and first perform a simple cut in proper motion-space to remove obvious non-members. Figure \ref{fig:pm_cut} shows the vector point diagram of all RGBs in the current sample. Red points in this figure show stars within 1 $-$ 2 half-light radii (r$_{h}$)  of NGC~5466, which form a tight clumping around the systemic proper motion of the cluster. We remove a large fraction of background contamination by only retaining stars whose proper motions fall within a 2 mas\,yr$^{-1}$ radius around the mean value for the cluster (red dashed circle). This radius is nearly 10 times larger than the average proper motion uncertainties of stars in the cluster’s core, and is therefore unlikely to remove any stars associated to the cluster. This cut greatly reduces the sample down to $\sim$6,600 stars. The top panel in Figure \ref{fig:Giants_Cuts} shows the tangent plane for all giants in the original dataset, while the second panel shows the RGBs remaining after the proper motion cut.

% \begin{figure}
%     \centering
%     \includegraphics[width=0.45\textwidth]{Figures/PM_Selection_Plot_nozoom.png}
%     % \includegraphics[width=0.15\textwidth]{Figures/PM_Selection_Plot_zoom.png}
%     \caption{Vector point diagram of all giants (P$_{giant}$ $>$ 50\%) in the CFIS dataset (see text for details). Stars within 1~$-$~2r$_h$ of the cluster are shown as red points. The red dashed circle is the 2~mas~yr$^{-1}$ boundary chosen to select stars with broadly similar proper motions to the globular cluster.}
%     \label{fig:pm_cut}
% \end{figure}

NGC~5466 has a metallicity of [Fe/H] $\simeq$ $-$2\,dex (\citealt[][2010 edition]{harris1996}) and no reported evidence of a spread in iron. Therefore, we require the metallicity for candidate stream members to be limited to a range of [Fe/H] = [$-$2.3, $-$1.7]\,dex. The range is set by  the global  uncertainty on the metallicity of the giants in our dataset  ($\pm$0.3 dex; see \citetalias{thomas2019}). The stars that remain after this cut are shown in the third panel of Figure~\ref{fig:Giants_Cuts}. 

Finally, we restricted the allowable heliocentric distance range for the RGB stars. This selection was made primarily based on two factors: firstly, NGC~5466 is located 16 kpc away, and secondly, typical distance uncertainties for our stars are $\sim$25\%. Further, the distance of stars along the stream are as of yet, unknown. It is quite possible that the putative stream may exhibit a significant distance gradient. For these reasons, we require R$_{helio}$ = [10, 22]~kpc. The remaining stars in our dataset are shown in the lower panel of Figure~\ref{fig:Giants_Cuts}.

\subsubsection{Identifying stream members}
\label{sect:sigclip}

\begin{figure*}
    \centering
    \includegraphics[width=0.4\textwidth]{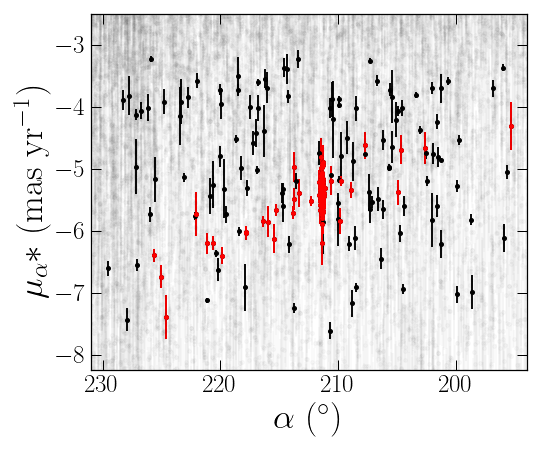}
    \includegraphics[width=0.41\textwidth]{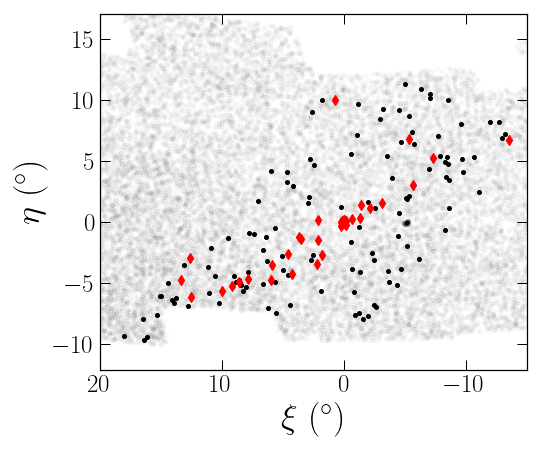}
    \caption{Left panel shows the proper motions (not corrected for Solar reflex motion) of RGB stars as a function of $\alpha$, not corrected for Solar motion. The right panel is the tangent plane of the same data. For both plots, grey points are the full RGB sample, black are the filtered sample, and red represent stars surviving the sigma-clipping routine.}
    \label{fig:sigma_clip}
\end{figure*}

The descending panels of Figure \ref{fig:Giants_Cuts} show tentative evidence of an extended structure south-east and north-west of NGC~5466. If this structure is real, then there should be a gradient in the proper motions of the relevant stars seen on-sky, i.e. a coherent phase-space structure. To investigate this possibility, we selected stars within a generous polygon defined by the green box in panel (d) of Figure \ref{fig:Giants_Cuts}. This selection is made to broadly cover the area of the putative stream, without being too restrictive. We then automatically search for any stars that are consistent with displaying a gradient in their proper motions. Specifically:

\begin{enumerate}
    \item We first examine the behaviour of $\mu_\alpha$* as a function of $\alpha$ for all stars. We fit a straight line to all the data, where the weights are given by the uncertainties in the individual proper motions;
    \item We retain those stars that are less then 3$\sigma$ from the fitted line. We then generate a new fit to these data;
    \item Using the fit from the previous step, we compare this fit to all the data (i.e. including stars that were rejected in the previous iteration), again only retaining stars less then 3$\sigma$ from the fitted line.
    \item We repeat step (iii) until convergence;
    \item For the stars that survive the sigma-clipping in $\mu_\alpha$* vs $\alpha$, we repeat the process for $\mu_\delta$ vs $\delta$.
    \end{enumerate}
    
Figure \ref{fig:sigma_clip} shows the impact of applying this procedure to the dataset. Red points in this figure are the surviving members of the sigma-clipping. It is notable that these form a clear, coherent, and extended stream-like structure on the sky as shown in the right panel, with only a few outliers. We note that we have repeated this analysis for different polygon shapes of the selection box in panel (d) of Figure \ref{fig:Giants_Cuts}, and we have verified that our conclusions are independent of the exact shape of the polygon. We conclude that the red data shown in Figure \ref{fig:sigma_clip} are a real detection of stars belonging to an extended stellar stream from NGC~5466. In what follows, we refer to these stars (beyond the tidal radius of NGC~5466) as our “gold” sample, and we additionally include the extra-tidal BHBs identified in Section \ref{sect:tidal_debris}. Properties of these stars are provided in Table \ref{tab:goldsample}.

% \begin{figure*}
%     \centering
%     % \includegraphics[width=0.45\textwidth]{Figures/SigmaClip_difference copy.png}
%     \includegraphics[width=0.4\textwidth]{Figures/SigmaClip_difference_new1.png}
%     \includegraphics[width=0.41\textwidth]{Figures/SigmaClip_difference_new2.png}
%     \caption{\textcolor{red}{Left panel shows the raw proper motions of RGB stars as a function of $\alpha$, not corrected for Solar motion. The right panel} is the tangent plane of the same data. For both plots, grey points are the full RGB sample, black are the filtered sample, and red represent stars surviving the sigma-clipping routine.}
%     \label{fig:sigma_clip}
% \end{figure*}

\section{Quantifying the properties of the stellar stream from NGC~5466}
\label{sect:4}

In this section, we use the gold sample to quantify the path of the stream and search for additional members. We then quantify the stream's morphology and estimate its luminosity.

\subsection{Defining a native coordinate system}

As a satellite orbits the Galaxy, its trajectory follows a path that is closely represented by the best fit great circle (\citealt{johnston1996,ibata2001}). For a relatively distant stream such as NGC~5466 that clearly spans several tens of degrees, a great circle fit in the heliocentric frame is a good approximation to the Galactocentric equivalent, and is highly convenient as a frame in which to quantify the stream's properties. Positions on the celestial sphere are given in ($\phi_1$, $\phi_2$) which describe the longitude and latitude, respectively, of the great circle. 

For the NGC~5466 reference frame, we determined the best fit pole of the great circle plane using stars from the gold sample. We used a least-squares minimization to minimize the scatter of stars in the $\phi_2$ coordinate such that the origin of this system is centered on the cluster. The resulting plane is defined by its pole at ($\alpha_{P}$,~$\delta_{P}$) = (-16.86 $\pm$ 0.83$^\circ$, 50.77 $\pm$ 0.46$^\circ$). The transformation from equatorial to stream coordinates is given by:

\begin{equation}
\begin{split}
    \begin{bmatrix}
        \cos(\phi_1) \cos(\phi_2) \\
        \sin(\phi_1) \cos(\phi_2) \\
        \sin(\phi_2) 
    \end{bmatrix} 
    = 
    R \times 
    \begin{bmatrix}
        \cos(\alpha) \cos(\delta) \\
        \sin(\alpha) \cos(\delta) \\
        \sin(\delta) 
    \end{bmatrix}
\end{split}
\label{eq:rotation}
\end{equation}

\noindent where the rotation matrix R is:

\begin{equation}
    R = 
    \begin{bmatrix}
        -0.7500 & -0.4572 & 0.4780  \\
        -0.2664 & 0.8702 & 0.4144  \\
        0.6054 & -0.1835 & 0.7745
    \end{bmatrix}
\end{equation}

% \begin{equation}
% \begin{split}
%     \begin{bmatrix}
%         \cos(\phi_1) \cos(\phi_2) \\
%         \sin(\phi_1) \cos(\phi_2) \\
%         \sin(\phi_2) 
%     \end{bmatrix} 
%     = \\
%     \begin{bmatrix}
%         -0.7500318816 & -0.4571694316 & 0.4779626423  \\
%         -0.2663830593 & 0.8702492573 & 0.414374584  \\
%         0.6053860275 & -0.1834729981 & 0.7744968797
%     \end{bmatrix}
%     \times \\
%     \begin{bmatrix}
%         \cos(\alpha) \cos(\delta) \\
%         \sin(\alpha) \cos(\delta) \\
%         \sin(\delta)
%     \end{bmatrix}
% \end{split}
% \label{eq:rotation}
% \end{equation}

\noindent Note that the x-axis must be inverted for the leading arm to correspond with increasing $\phi_1$. The south-east to north-west extent of the stream in this frame begins at ($\alpha$, $\delta$) $\simeq$ (198$\degr$, 36$\degr$) and ends near ($\alpha$, $\delta$) $\simeq$ (229$\degr$, 21$\degr$).

\begin{figure*}
    \includegraphics[width=1\textwidth]{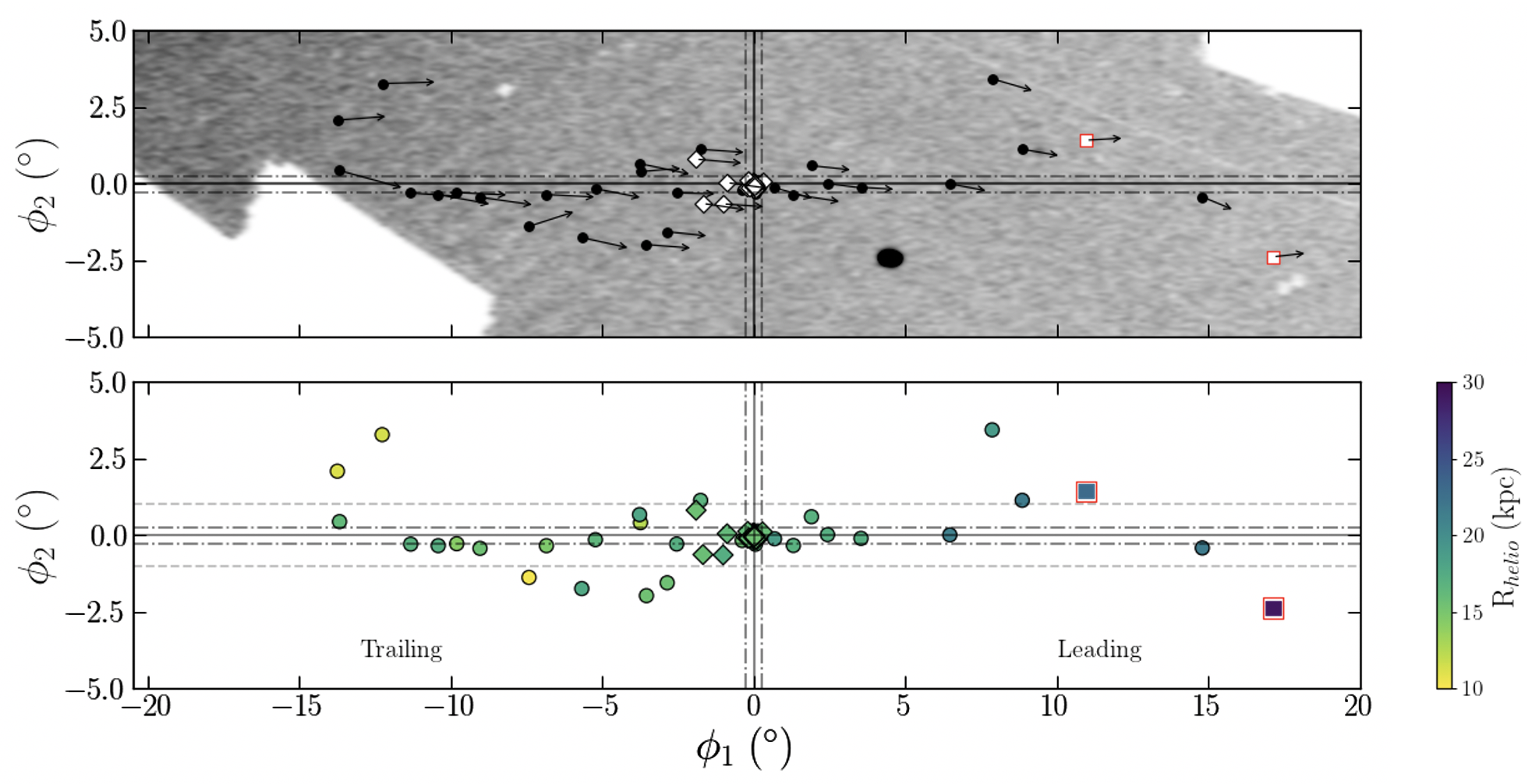}
    \caption{Gold sample RGBs (circles) and BHBs (diamonds) in the stream coordinate system. Vectors in the top panel are the proper motions corrected for Solar reflex motion, and rotated into the great circle frame. We show the limits of the CFIS footprint as the grey-filled region, where the dense circle at ($\phi_1$,~$\phi_2$)~$\sim$~(5$\degr$,~$-$2.5$\degr$) is the nearby cluster, M~3. Colors of stars in the bottom panel correspond to their heliocentric distances. Most of the stars identified are within 1$\degr$ of the $\phi_2$ plane, shown as the dashed lines in the bottom panel. Dashed-dotted lines signify $\pm$r$_{t}$ (tidal radius). The two additional member stars identified in Section \ref{sect:stream_kinematics} are plotted in red.}
    \label{fig:NGC5466_Phi_Gradients}
\end{figure*}

Figure \ref{fig:NGC5466_Phi_Gradients} shows the distribution of gold sample members (BHBs and RGBs as diamonds and circles, respectively) and giants within the cluster in this new frame of reference. The top panel shows the proper motion vectors for each star (scaled by 40\%), corrected for Solar reflex motion (as in Section \ref{sect:tidal_debris}) and rotated into the great circle frame. In the bottom panel, the color scale represents the heliocentric distances for each star.

%\subsection{Cluster Estimates}

%The complete sample of tracer stars associated with NGC~5466 consists of 197 giants (at present), 163 of which are RGBs and BHBs located within the tidal radius of the cluster. Prior to studying the stream in detail, we confirmed our measurements for the main body of the cluster to that of the literature. 

%The number density profile of NGC~5466 in the CFIS-PS1 dataset (including all stellar sources, and not solely the giants discussed here) shows crowding effects within a radius of 0.5 half-light radii (r$_{h}$). We recognize our estimates may be impacted by bias due to these crowding effects; therefore, we opt to only use the giants between r = [0.5, 2] r$_{h}$. Table \ref{tab:estimates} summarizes our findings and compares our mean quantities to the literature. We find good agreement with previous works.

\subsection{Searching for additional stream members}
\label{sect:stream_kinematics}

\begin{figure}
    \centering
    \includegraphics[width=0.45\textwidth]{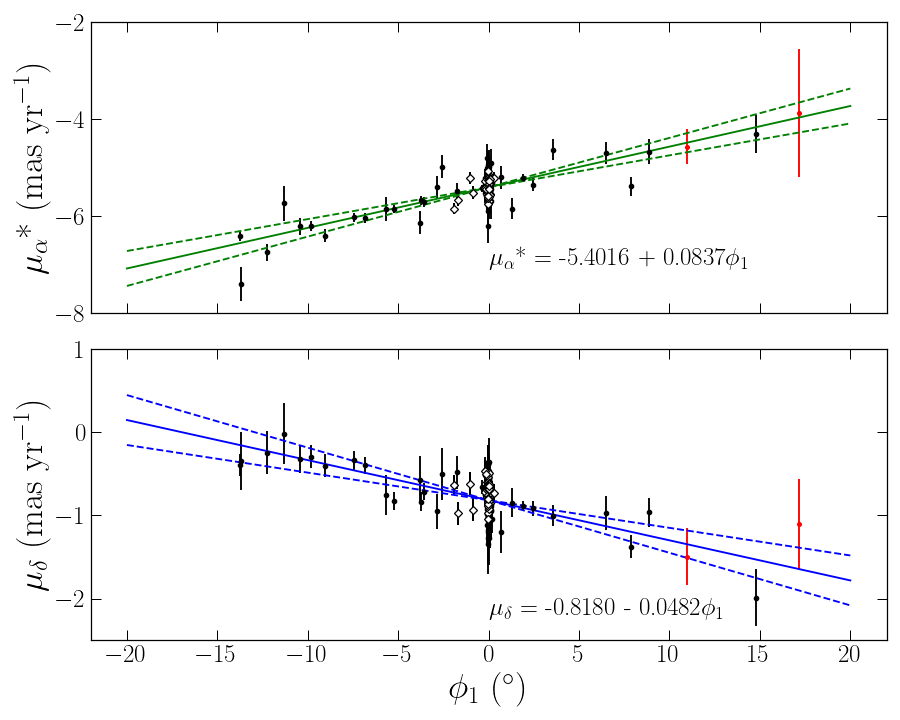}
    \caption{Proper motions, uncorrected for Solar reflex motion, as a function of $\phi_1$. Point styles are the same as in Figure~\ref{fig:NGC5466_Phi_Gradients} and additional members have been appended in red. The best fit trends are shown as solid lines. Dashed lines represent three times the standard error on the fitted slopes.}
    \label{fig:polynomials}
\end{figure}

Prior to quantifying the properties of the stream, we first search for any additional members that may have been missed during our initial cuts. The kinematics of the  gold sample stars are parameterized as a function of stream longitude. Specifically, we fit a polynomial to the proper motions in each direction as a function of $\phi_1$, and find a one-degree fit is favored over other models (as quantified via the Akaike Information Criterion, or AIC). Figure \ref{fig:polynomials} shows the best fit polynomials to the proper motion data in each direction, where the dashed lines show the 3$\sigma$ uncertainty in the slope of the fits. 

Using these parameterizations, we search the RGB and BHB datasets within the range $\phi_1$ = [-30$\degr$, 30$\degr$]  for any stars that have proper motions consistent with those of the stream. For the RGBs, we also require that the stars have consistent metallicities to that of the cluster (i.e., [Fe/H] = [$-$2.3, $-$1.7] dex). We consider a star a candidate stream member if the 1$\sigma$ range of its proper motion is consistent to that of the polynomial fits (to within three times the standard error of the fit, as shown by the dashed lines in Figure \ref{fig:polynomials}). An additional restriction required is that any putative members are within $|\phi_2|$ $\leq$ 5$\degr$. As inspection of Figure~\ref{fig:NGC5466_Phi_Gradients} makes clear, this is a generous cut, and only removes stars that are clearly far away from the plane of the stream.

% We consider the proper motions consistent for a given star if the 
% 1$\sigma$ range of its proper motion in consistent with the expected proper motion of the stream as given by the polynomial fits (to within three times the standard error of the fit, as shown by the dashed lines in Figure \ref{fig:polynomials}). 

Finally, we look at the heliocentric distances of the remaining stars and compare them to the trend of R$_{helio}$ vs $\phi_1$. We find two additional possible members using this method. These stars are appended to Table \ref{tab:goldsample} and we show them in Figures \ref{fig:NGC5466_Phi_Gradients} and \ref{fig:polynomials} in red.

\subsection{Quantifying the stream's properties}

\subsubsection{Length and width}

The maximum difference in longitude between the putative stream members is $\sim$31$\degr$, including the new member stars identified in the preceding section. We adopt this value as the length of (the visible part of) the stream. However, the stream's length may very well continue in the trailing arm ($\phi_1$ $<$ $-$15$\degr$), where the trajectory approaches the limits of the CFIS footprint. This also corresponds to the approximate bounds of the SDSS detection made by \citet{grillmair2006}. 

We estimate the stream's width by using the distribution of stars in $\phi_2$, beyond 1$\degr$ from the cluster center. We find the width dispersion of this stream is $\sigma_w$ = (1.31 $\pm$ 0.24)$\degr$, similar to the detected width found in \citet{grillmair2006}. In physical units, this is $\textit{w}$ = 367 $\pm$ 67 pc wide at the distance of NGC~5466. We note in Section \ref{sect:6} there are a handful of RGBs which may be considered outliers (see also bottom left panel in Figure \ref{fig:polygala}). After removing these stars, we obtain a similar width dispersion of $\sigma_w$ = (1.14 $\pm$ 0.22)$\degr$, or equivalently, $\textit{w}$ = 318 $\pm$ 61 pc wide.
% While this value is typically larger than most globular cluster streams (), 
% To measure the stream’s width, we fit a Gaussian to the $\phi_2$ distribution for all stream members beyond 1$\degr$ from the cluster center. We estimate the standard deviation of this fit is (0.90~$\pm$~0.11)$\degr$, or a FWHM~=~(2.13~$\pm$~0.27)$\degr$. This is approximately 3.5 times the tidal radius of NGC~5466. At a distance of 16 kpc, this is equivalent to a width dispersion of 250 $\pm$ 31 pc. We note in Section \ref{sect:6} there are some RGBs in our sample that may be considered outliers (see also Table \ref{fig:Giants_Cuts}); in removing these stars, we obtain a similar width dispersion of 229 $\pm$ 25 pc. 

The relative sparseness of stars in the leading arm ($\phi_1$ $>$ 0$\degr$) to that of the trailing arm should be noted. Although the lack of stars at $\phi_2$ $>$ 10$\degr$ could be caused by completeness effects, especially due to the large distances in the leading arm, in practice this is unlikely. The most distant RGB we find in the stream has a $\textit{Gaia G}$-band magnitude of 19.6, nearly 1.5 magnitudes brighter than the $\textit{Gaia}$ magnitude limit. We address a possible reason for this discrepancy in stream density, which could be due to the cluster's past interactions with the Galaxy, in Section \ref{sect:6}.

\subsubsection{Luminosity}

To determine a lower limit for the luminosity of the stream, we compare the number of giants identified within the tidal radius of NGC~5466 to those of the gold sample. We parameterize the cluster's density as a King profile and calculate the fraction of light in radial bins between [0.77$r_h$, r$_t$] using parameters from Table \ref{tab:ngc5466}. The lower limit of this range is set such that we avoid the central most regions of the cluster, where crowding effects in the CFIS data become significant. The fraction of light in this range is 29.5\% of the luminosity of the entire cluster. 

We estimate the number of giants (both RGBs and BHBs) between [0.77$r_h$, r$_t$] to be 129 $\pm$ 11. To estimate the total number of detectable giants, we extrapolate this profile and find  437~$\pm$~21 giants within the tidal radius. We compare this to the 36 giants we identify as stream members in this work, yielding a ratio of stars in the stream to those in the main body is 0.082 $\pm$ 0.004. 

NGC~5466's absolute visual magnitude is M$_V$ = $-$6.98 mag, which equates to a total luminosity of $\sim$4.9~$\times$~10$^{4}$\,L$_\odot$. This implies that the luminosity of the detected part of the stream is [4.0~$\pm$~0.2]~$\times$~10$^{3}$\,L$_\odot$. Adopting a stellar mass-to-light ratio of $\sim$1 (\citealt{pryor1991}) means the stellar mass in the stream is roughly 4.0~$\times$~10$^{3}$\,M$_\odot$.

% The stellar mass estimate of NGC~5466 from \citet{fellhauer2007} is 5~$\times$~10$^{4}$\,M$_\odot$ (and an estimated stellar mass-to-light ratio of $\sim$1 from \citealt{pryor1991}), meaning the stellar mass in the stream is roughly [4.1~$\pm$~0.2]~$\times$~10$^{3}$\,M$_\odot$. 

These estimations have a few caveats and should certainly only be considered as a lower limit. Most importantly, we only estimate the luminosity of the part of the stream that we have been able to detect. Lower surface brightness parts of the stream are not accounted for in this calculation, and of course we have no information on the stream beyond the CFIS footprint. Additionally, the stream detection is extremely sparse. For the 36 giants we identify in the stream, and given the length and width we estimated, this corresponds to a density of 0.55 $\pm$ 0.07 bright giants deg$^{-2}$.
%%%%%Recalculated
% Alternatively, this equates to an average surface brightness of \textbf{XXX} $\pm$ \textbf{XXX} mags arcsec$^{-2}$, given our stream luminosity measurement. 

% \begin{landscape}
% \begin{table}
% \renewcommand{\arraystretch}{0.75}
\begin{table*}
\caption{List of “Gold Sample” giants. IDs with an asterisk (*) are the possible contaminants identified in Figure \ref{fig:polygala}. The tentative final members added in Section \ref{sect:stream_kinematics} are marked with double (**).}
\centering
\label{tab:goldsample}
    \begin{tabular}{ccccccccccc}
    \hline
    No. & Population & $\alpha$ & $\delta$ & R$_{helio}$ & $\mu_{\alpha}$* &  $\mu_{\delta}$ & $\phi_1$ &  $\phi_2$ &  $\mu_{\phi_1}$ &  $\mu_{\phi_2}$ \\
    
    & & ($\degr$) & ($\degr$) & (kpc) & (mas\,yr$^{-1}$) & (mas\,yr$^{-1}$) & ($\degr$) & ($\degr$) & (mas\,yr$^{-1}$) & (mas\,yr$^{-1}$) \\
    \hline\hline
    % 1 & BHB  & 212.0302 & 27.5003 & 17.2967 $\pm$  1.7297 & -5.2112 $\pm$ 0.1305 & -0.6172 $\pm$ 0.1437 & & & & \\
    1 & BHB  &    212.0302 &    27.5003 & 17.2967 $\pm$  1.7297 & -5.2112 $\pm$ 0.1305 & -0.6172 $\pm$ 0.1437 & -1.0169 & -0.6473 &  3.5773 & -0.2422 \\
 
     2 & BHB  &    211.1383 &    28.7820 & 15.9393 $\pm$    1.5939 & -5.2123 $\pm$ 0.1166 & -0.7325 $\pm$ 0.1044 &  0.2826 &  0.1094 &  3.4694  & -0.0848 \\
     
     3 & BHB  &    212.2727 &    28.1715 & 16.2251 $\pm$    1.6225 & -5.5180 $\pm$ 0.1349 & -0.9305 $\pm$ 0.1404 & -0.8848 &  0.0448 &  3.6424  & -0.4972 \\
     
     4  & BHB &    212.6950 &    27.1924 & 15.8071 $\pm$    1.5807 & -5.6548 $\pm$ 0.1249 & -0.9769 $\pm$ 0.1363 & -1.6828 & -0.6347 &  3.7182  & -0.5320 \\
     
     5   & BHB &    213.7126 &    28.3461 & 15.5797 $\pm$    1.5580 & -5.8405 $\pm$ 0.1152 & -0.6391 $\pm$ 0.1208 & -1.9106 &  0.8110 &  4.0074  & -0.3401 \\
     
     6 & BHB & 211.6359 & 28.5638 & 16.0564 $\pm$ 1.6056 & -5.4169 $\pm$ 0.1140 & -0.4610 $\pm$ 0.1307 & -0.2053 & 0.1230 & 3.7772 & 0.0175 \\
     
     \hline
     
     7 & RGB   &    195.3068 &    34.2154 & 21.6520 $\pm$    1.3741 & -4.3092 $\pm$ 0.3881 & -1.9872 $\pm$ 0.3450 & 14.8034 & -0.4181 &  2.6937 & -1.1015 \\
 
     8 & RGB  &    202.6499 &    33.4567 & 21.6475 $\pm$    1.8703 & -4.6716 $\pm$ 0.2569 & -0.9615 $\pm$ 0.1753  &  8.8576 &  1.1352 &  3.2370 & -0.5511 \\
     
     9 & RGB  &    204.6825 &    31.4458 & 21.3098 $\pm$    1.6581 & -4.6976 $\pm$ 0.2340 & -0.9712 $\pm$ 0.2021  &  6.4674 &  0.0071 &  3.2133 & -0.5985 \\
     
     10  & RGB &    204.8897 &    35.1409 & 18.9778 $\pm$    2.0912 & -5.3810 $\pm$ 0.1986 & -1.3792 $\pm$ 0.1389 &  7.8654 &  3.4324 &  3.5611 & -1.0257\\
    
    11  & RGB &   207.7039 &    30.0776 & 17.2575 $\pm$    1.2324 & -4.6257 $\pm$ 0.2220 & -1.0034 $\pm$ 0.1258 &  3.5350 & -0.1058 &  2.9362 & -0.1681\\
    
    12  & RGB  &  208.9051 &    29.6896 & 17.2590 $\pm$    1.2863 & -5.3480 $\pm$ 0.1310 & -0.9136 $\pm$ 0.0914 &  2.4302 &  0.0147 &  3.6020 & -0.4556 \\
    
    13 & RGB   &  209.8709 &    28.8610 & 16.9963 $\pm$    0.8251 & -5.8456 $\pm$ 0.2130 & -0.8476 $\pm$ 0.1748 &  1.3021 & -0.3374 &  4.0484 & -0.6151\\
    
    14 & RGB  &    209.7578 &    29.9678 & 17.0305 $\pm$    0.8117 & -5.2011 $\pm$ 0.0765 & -0.8851 $\pm$ 0.0542 &  1.8989 &  0.5999 &  3.4566 & -0.3642 \\
    
    15 & RGB &    211.1529 &    28.3138 & 16.4268 $\pm$    0.4823 & -5.3972 $\pm$ 0.0669 & -0.9246 $\pm$ 0.0661 &  0.0512 & -0.2978 &  3.5681 & -0.4263 \\
    
    16  & RGB &    210.6149 &    28.7657 & 18.6889 $\pm$    1.0345 & -5.1982 $\pm$ 0.2352 & -1.2038 $\pm$ 0.2508 &  0.6804 & -0.1198 &  3.3623 & -0.9172 \\
    
    17 & RGB  &    211.6564 &    28.2156 & 15.5932 $\pm$    0.7845 & -5.4239 $\pm$ 0.0779 & -0.6604 $\pm$ 0.0823 & -0.3861 & -0.1750 &  3.6677 & -0.0695 \\
    
    18  & RGB &   213.3529 &    25.8162 & 15.6223 $\pm$    0.9959 & -5.3910 $\pm$ 0.2285 & -0.9500 $\pm$ 0.2119 & -2.8641 & -1.5543 &  3.4879 & -0.3571 \\
    
    19 & RGB  &    213.7877 &    25.1158 & 15.5186 $\pm$    0.5563 & -5.7101 $\pm$ 0.1064 & -0.7134 $\pm$ 0.0967 & -3.5489 & -1.9741 &  3.8738 & -0.2963 \\
    
    20 & RGB  &    213.7368 &    27.0737 & 17.2229 $\pm$    0.8812 & -4.9750 $\pm$ 0.2347 & -0.5004 $\pm$ 0.3135 & -2.5511 & -0.2887 &  3.4015 & -0.0761 \\
    
    21  & RGB &    213.7444 &    28.6991 & 16.8205 $\pm$    1.1477 & -5.4889 $\pm$ 0.1831 & -0.4822 $\pm$ 0.1947 & -1.7624 &  1.1325 &  3.8318 & -0.2637 \\
    
    22 & RGB  &    215.2901 &    27.0877 & 12.4913 $\pm$    0.4596 & -5.6743 $\pm$ 0.0975 & -0.8364 $\pm$ 0.1119 & -3.7466 &  0.4070 &  3.5615 &  0.2789  \\
    
    23 & RGB  &    215.4653 &    27.2993 & 18.0070 $\pm$    1.0454 & -6.1283 $\pm$ 0.2307 & -0.5721 $\pm$ 0.2842 & -3.7757 &  0.6681 &  4.3691 & -0.9047 \\
    
    24 & RGB  &    215.9519 &    24.2550 & 17.7626 $\pm$    1.8410 & -5.8570 $\pm$ 0.2503 & -0.7543 $\pm$ 0.2383 & -5.6841 & -1.7440 &  4.0432 & -0.8798 \\
    
    25 & RGB  &    216.4128 &    25.8587 & 16.4943 $\pm$    0.7597 & -5.8503 $\pm$ 0.0857 & -0.8306 $\pm$ 0.1060  & -5.2335 & -0.1493 &  3.9321 & -0.7682  \\
    
    26*  & RGB &    217.7949 &    23.6799 & 10.3904 $\pm$    0.4976 & -6.0207 $\pm$ 0.0979 & -0.3381 $\pm$ 0.1120 & -7.4265 & -1.3794 &  3.9355 &  1.2545 \\
    
    27 & RGB   &    217.8405 &    24.8631 & 14.7913 $\pm$    0.9061 & -6.0378 $\pm$ 0.1080 & -0.3962 $\pm$ 0.1034 & -6.8513 & -0.3446 &  4.2221 & -0.2042 \\
    
    28 & RGB  &    219.8409 &    23.6443 & 15.5834 $\pm$    0.8387 & -6.4060 $\pm$ 0.1342 & -0.4024 $\pm$ 0.1373 & -9.0432 & -0.4306 &  4.5507 & -0.6402 \\
    
    29 & RGB  &    221.1630 &    22.9827 & 17.2438 $\pm$    0.7690 & -6.2060 $\pm$ 0.1721 & -0.3184 $\pm$ 0.1686 &    -10.4232 & -0.3438 &  4.4759 & -0.8088 \\
    
    30 & RGB  &    220.6388 &    23.3639 & 14.3005 $\pm$    0.6498 & -6.2007 $\pm$ 0.1058 & -0.2952 $\pm$ 0.1365 & -9.8121 & -0.2797 &  4.3653 & -0.1946 \\
    
    31 & RGB  &    222.0228 &    22.5356 & 16.9056 $\pm$    1.3742 & -5.7303 $\pm$ 0.3589 & -0.0170 $\pm$ 0.3709 &    -11.3320 & -0.2929 &  4.2191 & -0.2725 \\
    
    32 & RGB  &    224.5786 &    21.8593 & 16.3049 $\pm$    1.0889 & -7.3919 $\pm$ 0.3478 & -0.3471 $\pm$ 0.3517 &    -13.6814 &  0.4400 &  5.3777 & -1.4567 \\
    
    33*  & RGB &    225.0265 &    24.9984 & 11.4767 $\pm$    0.8607 & -6.7452 $\pm$ 0.1837 & -0.2462 $\pm$ 0.2571 &    -12.2730 &  3.2758 &  4.6083 &  0.1346  \\
    
    34*  & RGB &    225.6367 &    23.1840 & 11.2105 $\pm$    0.4497 & -6.4005 $\pm$ 0.1028 & -0.3936 $\pm$ 0.1299 &    -13.7523 &  2.0848 &  4.2402 &  0.3159 \\
    % \hline\hline\hline
    
    35** & RGB& 191.8986 & 33.1122 & 29.1899 $\pm$ 3.0057 & -3.8701 $\pm$ 1.3128 & -1.1017 $\pm$ 0.5421 & 17.1370 & -2.3727 & 2.8652 & 0.3140 \\
        
    36** & RGB& 200.4530 & 34.5566 & 23.0920 $\pm$ 1.7470 & -4.5673 $\pm$ 0.3618 & -1.4991 $\pm$ 0.3415 &   10.9650 &    1.4306 & 3.1950 &  0.1719 \\
    
    \hline
    \end{tabular}
\end{table*}
% \end{table}
% \end{landscape}

\section{Modelling the Dynamics of NGC~5466}
\label{sect:5}

Given the difficulty of earlier work in matching the path of the stream as defined by \cite{grillmair2006}, we now examine some simple dynamical models to aid in our understanding of NGC~5466’s orbital history. We consider both a point-mass orbit and a particle-spray model (see \citealt{fardal2015} for details) which we implement using the \textsc{python}-wrapped package known as \textsc{gala} (\citealt{pricewhelan2017}).

We use a right-handed Galactocentric coordinate system such that the Sun is located at (X, Y, Z) = ($-$8.122, 0.0, 0.025) kpc, with LSR velocities of [U,V,W]$_{\odot}$ = [11.1, 12.24, 7.25] km\,s$^{-1}$ (as in Section \ref{sect:tidal_debris}). The Milky Way potential is fixed using a three-component model, consisting of a \citet{miyamoto1975} disk, a \citet{hernquist1990} bulge, and spherical NFW dark matter halo (\citealt{navarro1996}):

\begin{equation}
    \Phi_{bulge}(R) = \frac{-GM_{b}}{R + a} %-  \frac{GM_{nucl}}{R + a_2}
    \label{eq:bulge}
\end{equation}

\begin{equation}
    \Phi_{disk}(R_{cyl}, z) = \frac{-GM_{d}}{\sqrt{R_{cyl}^2 + (b + \sqrt{z^2 + c^2})^2}}
    \label{eq:disk}
\end{equation}

\begin{equation}
    \Phi_{halo}(R) = \frac{-4 \pi G \rho_s r_s^3}{R} \ln{(1 + \frac{R}{r_s})}
    \label{eq:nfw}
\end{equation}

\begin{table}
    \centering
    \caption{Galactic potential parameters.}
    % \caption{Galactic potential parameters used in Equations \ref{eq:bulge}, \ref{eq:disk}, and \ref{eq:nfw}.}
    \begin{tabular}{c|c|c}
        \hline
        Component & Parameter & Value \\
        \hline\hline
        Bulge & M$_b$ & 5 $\times$ 10$^9$ M$_{\odot}$ \\
         & a & 1 kpc \\
        % M_{nucl} & 2 x 10^8 M_\odot \\
        % a_2 & 100 pc \\
        \hline
        Disk & M$_d$ & 5.56 $\times$ 10$^{10}$ M$_{\odot}$ \\
         & b & 3.5 kpc \\
         & c & 280 pc \\
        \hline
        Halo & M$_{vir}$ & 8.2 $\times$ 10$^{11}$ M$_{\odot}$ \\
         & R$_{vir}$ & 207 kpc \\
         & c$_{h}$ & 12 \\
         & r$_s$ & 17.25 kpc \\
        \hline
    \end{tabular}
    \label{tab:potential}
\end{table}

\noindent where R is the Galactocentric radius, R$_{cyl}$ is the cylindrical radius, and z is the vertical height above the disk. 

Our chosen values for masses, scale radii, and other parameters are summarized in Table \ref{tab:potential}. For the disk, the mass (M$_{d}$), scale height (b), and scale length (c) are the values used in the MWPotential2014 from \citet{bovy2015}. For the halo, we adopt recent estimates for the Milky Way virial mass and radius (M$_{vir}$ and R$_{vir}$, respectively) from \citet{cautun2020}. Assuming a dark matter concentration of 12 (a typical value observed in simulated Milky Way analogues; see \citealt{boylankolchin2010}) and using the relation R$_{vir}$ = c$_{h}$R$_{s}$, we adopt a scale height for the dark matter halo of 17.25 kpc. These parameters produce a circular velocity at the Sun similar to the estimate from \citet{eilers2019} (v$_{circ}$(R$_{\odot}$) = 229 km\,s$^{-1}$).

For our particle-spraying simulations, we account for the internal gravity of the satellite, represented as a spherical  \citet{plummer1911} profile:

\begin{equation}
    \Phi_p(r) = \frac{-GM_{sat}}{\sqrt{r^2 + d^2}}
    \label{eq:plummer}
\end{equation}

\noindent We used the present-day stellar mass of NGC~5466 and a scale length (d) of 6.7 pc derived from the core radius of the cluster (see Table \ref{tab:ngc5466}).

The point-mass orbit of NGC~5466 was integrated forwards and backwards in time by 1 Gyr to examine the trajectory of the cluster in our chosen Galactic potential. This serves as a rough estimate to the most likely path of the stream. 

For the particle-spray method, we initiated the cluster 2 Gyrs ago and ran the model forward in time, releasing one particle at each Lagrange point per timestep (for a total of 1,000 timesteps every 100 Myrs). Mass loss is not explicitly tracked in this method, therefore saving on computation time; however, each ejected particle is provided orbit information based on the current position and velocity of the progenitor and integrated forward with the cluster, allowing the stream's evolution to be mapped in phase-space. We examine the qualitative resemblances between the observational data and dynamical model in the following section.

\section{Discussion}
\label{sect:6}

\begin{figure}
    \centering
    \includegraphics[width=0.32\textwidth]{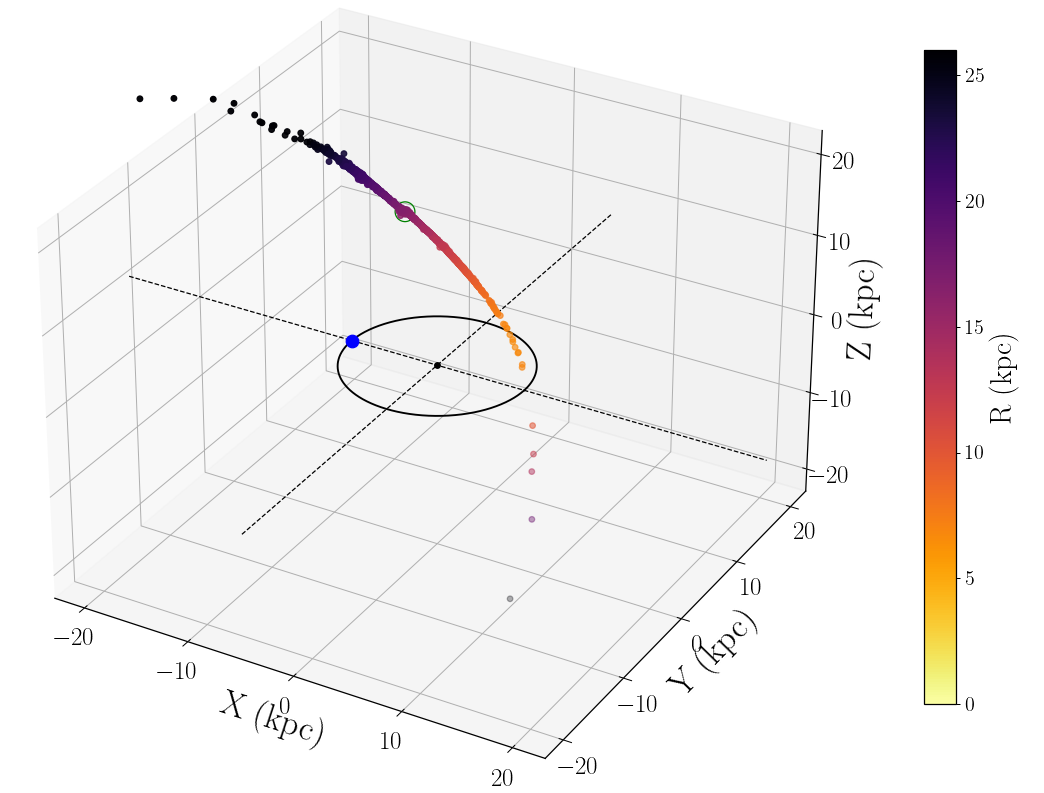}
    \includegraphics[width=0.29\columnwidth]{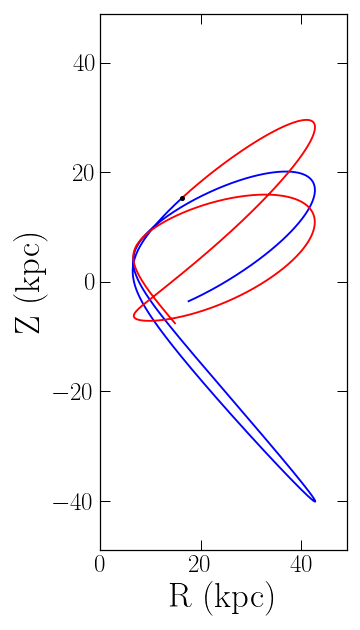} 
    \caption{Orbit integrations of NGC~5466 in the Galactic frame. The left plot shows a 3D representation of the sprayed particles, where the scaled colors correspond to Galactocentric radius (R). We also show the Solar circle where the Sun is located at the blue point. In the right panel, the blue and red trajectories are the backwards and forwards point-mass orbit integrations of NGC~5466.}
    \label{fig:orbit}
\end{figure}

We show the point-mass trajectory of the cluster in Figure \ref{fig:orbit}, where the backwards and forwards orbit integrations are represented as blue and red lines, respectively. From this estimation, we find that the pericenter is approximately 6.42 kpc, apocenter is 42.99~kpc, and the eccentricity of NGC~5466's orbit is 0.74. These values agree closely with those estimated in \citet{baumgardt2019}.

We note two significant points from the (R, z) panel on the right in Figure \ref{fig:orbit}: firstly, the cluster’s most recent pericentric passage occurred $\sim$50 Myrs ago, and secondly, NGC~5466 recently crossed the Galactic disk. Both interactions (especially at nearly the same time) suggest significant and recent mass loss. Previous simulations modeling the cluster's detailed disruption have obtained similar conclusions, claiming that NGC~5466 has suffered disk-shocking (\citealt{odenkirchen2004,fellhauer2007}). In particular, \citet{fellhauer2007} used $\textsc{superbox}$ particle-mesh simulations fit to the stream’s surface density and width based on the matched filter path given by the \citet{grillmair2006} detection, and present-day cluster measurements given by $\textsc{Hipparcos}$ and \citet{harris1996}. Their work suggested that the tidal tails of NGC~5466 could be as lengthy as 100$\degr$ on the sky $-$ for comparison, this is nearly the same order as the length of the Sagittarius stream (\citealt{ibata2002,majewski2003}). Though it would be worthwhile to calculate an updated, post-$\textit{Gaia}$ estimate for the cluster's disruption, we consider it beyond the present scope of this work. We leave it to future efforts to utilize $\textit{Gaia}$ astrometry with more detailed simulations matched to the gold sample.

\begin{figure*}
    \centering
    \includegraphics[width=1\textwidth]{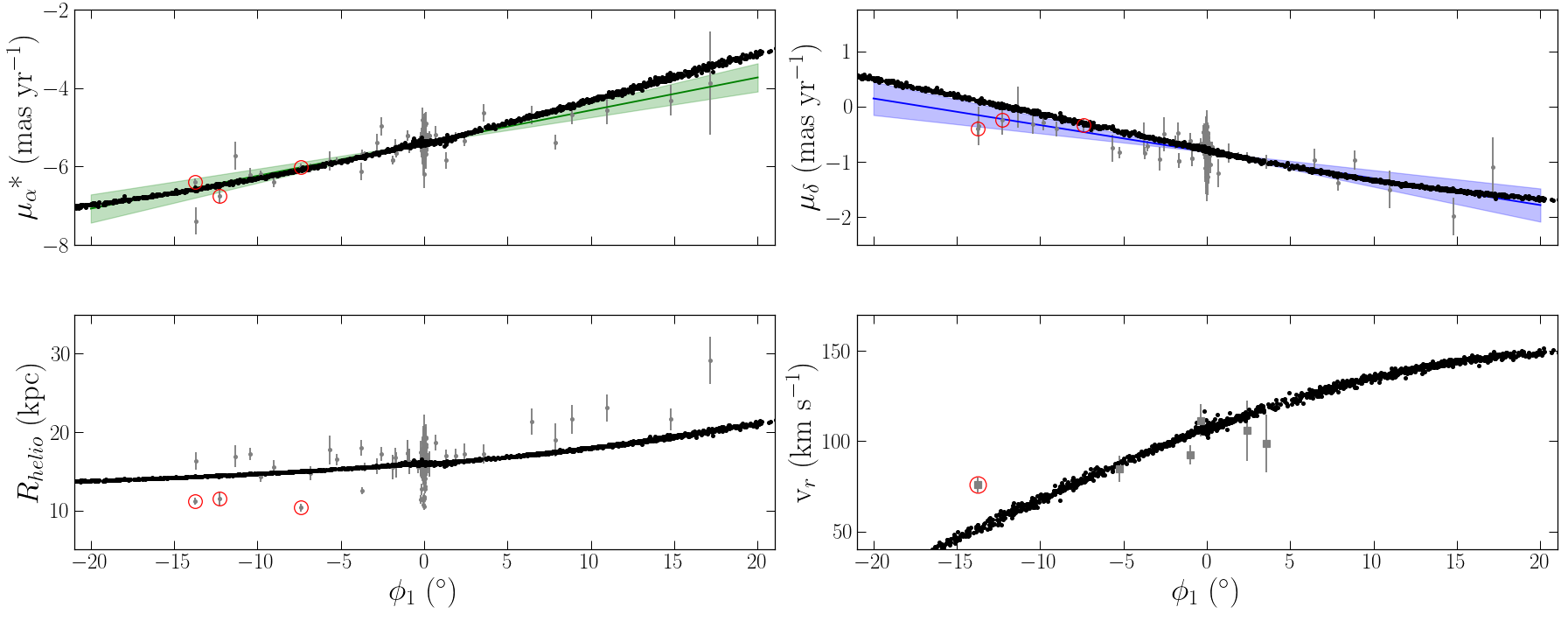}
    \caption{Comparison of kinematics against $\phi_1$ in the gold sample stars (grey error bars) and sprayed particles from the dynamical model (black points). The top panels show uncorrected proper motions in $\alpha$ and $\delta$, where the highlighted ranges show the upper and lower bounds of the fitted lines (same as in Figure \ref{fig:polynomials}). We show the heliocentric distances in the bottom left panel. Based on their discrepant distances, three gold sample members are highlighted in each panel with red circles. The final panel in the bottom right show the line-of-sight velocities for six stars present in LAMOST. Note that one particular outlier at ($\phi_1$, R$_{helio}$) = (-13.7$\degr$, 11.2 kpc) corresponds to the red circle in the final panel, which further confirms this star is an outlier based on its inconsistent radial velocity.}
    \label{fig:polygala}
\end{figure*}

Though the purpose of the particle-spray model was to compare the kinematic trends of star particles to our gold sample (and not to model the disruption of NGC~5466 over time), we find good agreement between observations and our simple model. Figure \ref{fig:polygala} shows the proper motions and heliocentric distances as functions of $\phi_1$, where the particles are represented as black points and the gold sample members are shown as grey error bars. We find that the global behavior of the proper motions is well-replicated by the model as a function of longitude. However, our simplistic model appears to systematically underestimate the distances in the leading arm (although there are not as many identified points at these larger distances). It is quite possible that this divergence may be a result of the spherical dark matter halo shape we implemented; alternatively, it could perhaps be due to an interaction with another satellite. Our preliminary analysis suggests NGC~5466 may have passed close to the LMC in its recent trajectory; however, this requires more investigation with a time-dependent potential. 

There are three gold sample members in Figure \ref{fig:polygala} whose proper motions are consistent with the rest of the data and the model, but whose distances do not follow the global trends. We consider these three points, highlighted with red circles, as outliers/contamination, and they are noted in Table~\ref{tab:goldsample} with an asterisk. We note that these additional members do not greatly affect estimates for the stream width, mass, or length. 

Lastly, we cross-matched the gold sample to spectra available from LAMOST (Large sky Area Multi-Object Fiber Spectroscopic Telescope; \citealt{luo2015}) and obtained radial velocities for six gold sample members. We show the comparison between in LAMOST DR6 radial velocities to $\textsc{gala}$ particles, which indeed appear consistent for all except one star that has already been labeled a contaminant based on its heliocentric distance. Summarized observational data for these stars are located in Table \ref{tab:LAMOST}.

\begin{table}
    \centering
    \caption{Six cross-matched stars observed in LAMOST DR6. Numbers in first column associate these stars to the data in Table \ref{tab:goldsample}. The star with (*) is highlighted as a contaminant based on its discrepant distance.}
    \begin{tabular}{ccccc}
        \hline
        No. & LAMOST ID &   $\alpha$ &  $\delta$ & v$_{r}$ \\
        & & ($\degr$) & ($\degr$) & (km\,s$^{-1}$) \\
        %  Parameter & Value & Source \\
         \hline\hline
            1 & J140807.25+273001.2 & 212.0302 &  27.5003 & 92.6 $\pm$ 5.4 \\
            11 & J135048.93+300439.3 & 207.7039 &  30.0776 & 98.8 $\pm$ 15.9 \\
            12 & J135537.21+294122.9 & 208.9051 &  29.6896 & 105.9 $\pm$ 16.7 \\
            17 & J140637.54+281256.3 & 211.6564 &  28.2156 & 111.4 $\pm$ 9.0 \\
            %   21 & J141458.68+284156.5 & 213.7444 &  28.6991 & -2987701200.0000 $\pm$ -2987701200.0000 \\
            25 & J141458.68+284156.5 & 216.4128 &  25.8587 & 84.6 $\pm$ 7.3 \\
            34* & J150232.78+231102.6 & 225.6367 &  23.1840 & 75.8 $\pm$ 4.2 \\
         
         \hline
    \end{tabular}
    \label{tab:LAMOST}
\end{table}

% Our ultimate goal is to present all kinematic tracers identified using the methods in this work

Given that we find broad agreement between data and model, we also compare to previous simulations of NGC~5466. Two works in particular found opposing conclusions, though both relied on the observed stream path given by the matched filter of \cite{grillmair2006}. Critically, \citet{fellhauer2007} found that it was not possible to reproduce the stream with the current proper motion estimate for the globular cluster. At the time of their study, the proper motions were derived from $\textsc{Hipparcos}$, which yielded ($\mu_{\alpha}*$, $\mu_{\delta}$) = ($-$4.65 $\pm$ 0.82, +0.8 $\pm$ 0.82) mas\,yr$^{-1}$. The authors opted for a lower value of $\mu_{\delta}$ = 0.4 mas\,yr$^{-1}$ (within 1$\sigma$ of the measurement) to best fit the tails. Compared to the more recent $\textit{Gaia}$-based estimate derived by \citet{baumgardt2019}, $\mu_{\delta}$ = $-$0.79 $\pm$ 0.01, their correction is closer to the more recent estimate. Overall, \citet{fellhauer2007} predicted a trend in heliocentric distance that is generally consistent with our new findings.

\citet{lux2012} developed a second dynamical model of NGC~5466 using the orbit-fitting method to constrain the Galactic potential. Interestingly, they reproduced the path of the stream up to $\alpha$ = 192$\degr$ (we note that it is at $\sim$195$\degr$ where we no longer detect any stream members) using oblate and triaxial halo shapes and claimed that spherical and prolate dark matter halos could be rejected at high confidence. However, the gradient in heliocentric distances predicted by \citet{lux2012} demonstrates the opposite trend of both \citet{fellhauer2007} and our new observations, such that stars in their model are at close distances of $\sim$10 kpc in the leading arm ($\phi_1$ $>$ 15$\degr$; see Figure 3 in \citealt{lux2012}). We note that, at the time of these studies, there was no direct information on the distance gradient or proper motions of stars in the stream for comparison. 

\begin{figure*}
    \centering
    \includegraphics[width=1\textwidth]{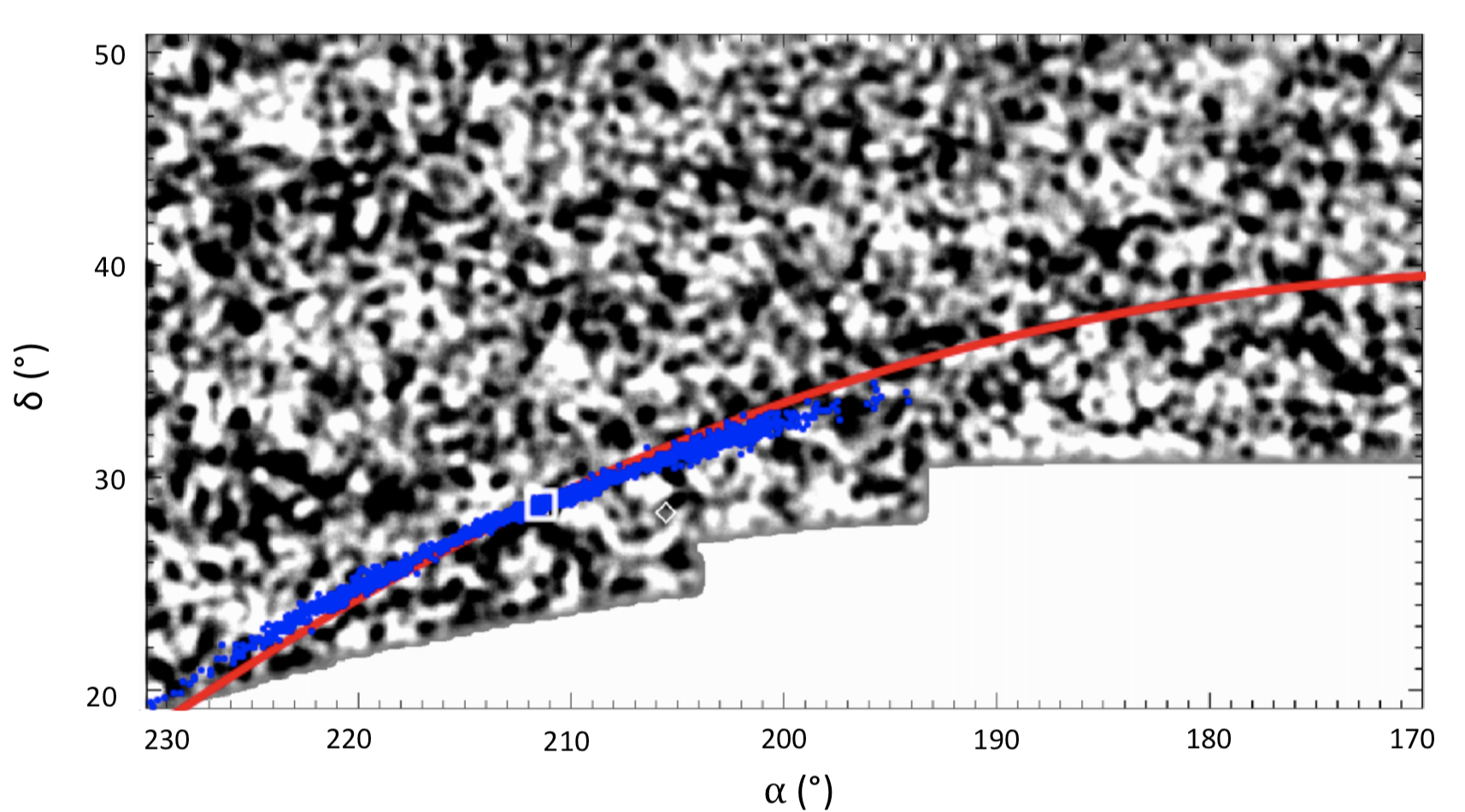}
    \caption{A direct comparison between the 45$\degr$ detection from \citet{grillmair2006} to this work. The path of the extended tails is claimed to begin at ($\alpha$, $\delta$) = (230$\degr$, 20$\degr$) and end near (180$\degr$, 42$\degr$). We overlay the great circle longitude ($\phi_1$; red line) and $\textsc{gala}$ star particles (blue) on the matched filter map. Both diverge at approximately $\alpha$ = 200$\degr$ compared to the original detection. NGC~5466 is represented with a white square and the nearby globular cluster M~3 (NGC~5272) as the diamond.}
    \label{fig:grillmair}
\end{figure*}

It is worthwhile to examine the position of the stream as detected in CFIS to the matched filter detection of \citet{grillmair2006}, which we show in Figure \ref{fig:grillmair}. \citet{grillmair2006} proposed that the dark diagonal strip from bottom left to top right corners of the plot is a stream from NGC~5466. We plot our great circle plane fit to the gold sample stars as a red line for comparison, and additionally highlight $\textsc{gala}$ star particles in blue. Close to the cluster, our findings broadly agree with \citet{grillmair2006}. However, at $\alpha$ $\lesssim$ 200$\degr$, the trajectories of our great circle and model deviate from the claimed matched filter path, and indeed, we are unable to detect many stream members past this position. Given our analysis is based on deeper photometry (CFIS and PS1~3$\pi$, compared to SDSS), we consider it unlikely that the feature in the matched filter at these coordinates is a real signal. This would also serve to explain previous difficulties in matching models to these observations.

In this work, we kinematically detect the tidal tails of NGC~5466 despite the fact that the stream is a) extremely diffuse and b) very distant, such that $\textit{Gaia}$ DR2 parallaxes are of limited use. We attribute our success to the tracer populations which have been robustly identified with CFIS, such that we obtain excellent distance estimates for stars at large Galactocentric radius. Paired with the exquisite proper motions in $\textit{Gaia}$ DR2, we are able to compile a detailed dataset of stars to explore the kinematic trends of this interesting stream. Though there have been many data-mining efforts implemented to probe $\textit{Gaia}$ DR2 for substructures (e.g., \citealt{mateu2018,ibata2019,necib2020,borsato2020}), our study is the only one to observe member stars of the NGC~5466 stellar stream within this dataset. Our work predated the arrival of $\textit{Gaia}$ eDR3, however, a recent study by \citet{ibata2020} also identified some stars in the NGC~5466 stream using the $\textsc{streamfinder}$ algorithm (\citealt{malhanibata2018}) applied to the newest $\textit{Gaia}$ release. In comparison, their detection identified mostly stars in the trailing arm and over a smaller angle on the sky than presented here.

% This work predated the arrival of $\textit{Gaia}$ EDR3, and a recent analysis from \citet{ibata2020} using the $\textsc{streamfinder}$ algorithm (\citealt{malhanibata2018}) applied to EDR3 successfully identified some stars in the NGC~5466 stream (although mostly only in the trailing arm, and over a smaller angle on the sky than presented here). 

%\begin{figure*}
%    \centering
%    \includegraphics[width=\textwidth]{Figures/NGC5466_equatorial.png}
%    \caption{Gold Sample stream in equatorial (top) and in the rotated great circle plane (bottom) where RGBs and BHBs are as in Figure \ref{fig:NGC5466_Phi_Gradients}. We searched over all grey points in this representation (up to $\phi_1$ = $\pm$30$\degr$) for additional members; these are therefore the RGBs limited by only metallicity. Red points are the original Gold Sample, and red/white squares show the appended members. \textbf{might remove => did not introduce figure in text}}
%    \label{fig:stream_eq}
%\end{figure*}

\section{Summary}
\label{sect:7}

In this work, we explored substructure in the outer stellar halo using multiple tracer populations with CFIS. We first examined the spatial distribution of CFIS BHBs with OPTICS and identified several known satellites  within the dataset, all of which showed some evidence of spatially extended stellar populations. Among these was the globular cluster NGC~5466, which is a distant cluster that had previously been argued to possess a long stellar stream, although no individual stellar members of the stream had been identified. 

Given its potential use to help probe the dynamics of the outer halo, and given that previous dynamical studies had difficulties reproducing the reported properties of the stream, we chose to further explore this system. We confirmed association of several of the extra-tidal BHBs  using cross-matched $\textit{Gaia}$ DR2 proper motions. By mining the expansive CFIS RGB dataset for stars whose kinematics and metallicities broadly represented that of the cluster, we found an extended stream from NGC~5466 that is both spatially and kinematically coherent.

Having identified stream members, we determined the stream's natural frame of reference and quantified its overall structure, dynamics, and stellar mass. We compared the observed behavior in proper motions and distances to simple dynamical models involving both a point mass and a $\textsc{gala}$ particle-spray model. We found that even these simple models are able to reproduce the global behavior of the stream's proper motion, and provide a good match to the observed distance gradient. Our work is the first to identify member stars of NGC~5466 both spatially and dynamically (predating $\textit{Gaia}$ eDR3), and we identify interesting systematic difference between observations and the models in the leading arm of the stream. 

We anticipate these results will motivate future modeling and observing campaigns (e.g., radial velocities), which could provide some interesting insights into the global shape of the Milky Way halo at these large distances. Our $\textsc{gala}$ model implements a spherical NFW halo and successfully reproduces, to first order at least, the major trends we observe in proper motions and distances. Unlike the previous claim by \cite{lux2012} based on earlier observations (which we show did not correctly trace the stream's path), it appears that the NGC~5466 stream cannot yet rule out a spherical halo shape. While NGC~5466 is only one stream out of a multitude of such structures, we expect it to be a very useful laboratory for those seeking to better measure the mass and shape of the Milky Way's gravitational potential. The fact that NGC~5466 is so distant, exhibits a strong distance gradient, has an identified progenitor, and is believed to have recently had an interaction with the Galaxy's disk, makes it a unique test-case for dynamical modelling of the Milky Way.

In the context of observational constraints on NGC~5466, we anticipate that the dynamical properties of the stream quantified in Section \ref{sect:stream_kinematics} will help in identifying additional member stars. This is a particular priority  for the trailing arm, which could well extend beyond the SDSS and CFIS footprints. The other obvious observational constraint currently lacking from our analysis are the radial velocities for member stars. In this era of $\textit{Gaia}$, it is increasingly the case that tangential velocities are more readily available than radial velocities, the exact opposite of what it has been for many decades. However, obtaining the stream's full 6-D kinematics will allow us to explore the phase-space distribution in detail $-$ including deriving energies and angular momenta for our member stars. With spectra of sufficiently high resolution, we can also better explore the chemical abundances of our stream sample. Already, this type of analysis on the main body has shown trends in $\alpha$-elements similarly observed in dwarf galaxies (\citealt{venn2004,lamb2015}), and follow-up spectra of [$\alpha$/Fe] abundances along the NGC~5466 stream should soon be possible with WEAVE (\citealt{dalton2012}).

It is notable that this stream was identified in $\textit{Gaia}$ DR2 when combined with photometric parallaxes from CFIS, that currently do not cover the full intended footprint of this survey. Using the all-sky power of $\textit{Gaia}$, and the wide field $\textit{u}$-band perspective of CFIS for 10,000 deg$^{2}$ in the north, soon to be followed by the Legacy Survey of Space and Time (LSST at the Vera C. Rubin Observatory) in the south, our exploration of the outer Galaxy is only just beginning.
% Large Synoptic Survey Telescope?? Guillaume says wrong acronym

\section*{Acknowledgements}

GT acknowledge support from the Agencia Estatal de Investigaci\'on (AEI) of the Ministerio de Ciencia e Innovaci\'on (MCINN) under grant with reference (FJC2018-037323-I). RI acknowledges funding from the European Research Council (ERC) under the European Unions Horizon 2020 research and innovation programme (grant agreement No. 834148). NFM and RI gratefully acknowledge support from the French National Research Agency (ANR) funded project ``Pristine'' (ANR-18-CE31-0017)

The work detailed above was conducted at the University of Victoria in Victoria, British Columbia, as well as in the Township of Esquimalt in Greater Victoria. We acknowledge with respect the Lekwungen peoples on whose unceded traditional territory the university stands, and the Songhees, Esquimalt and WS\'{A}NE\'{C} peoples whose historical relationships with the land continue to this day. 

CFIS is conducted at the Canada-France-Hawaii Telescope on Maunakea in Hawaii. We also recognize and acknowledge with respect the cultural importance of the summit of Maunakea to a broad cross section of the Native Hawaiian community.

This work is based on data obtained as part of the Canada-France Imaging Survey (CFIS), a CFHT large program of the National Research Council of Canada and the French Centre National de la Recherche Scientifique. Based on observations obtained with MegaPrime/MegaCam, a joint project of CFHT and CEA Saclay, at the Canada-France-Hawaii Telescope (CFHT) which is operated by the National Research Council (NRC) of Canada, the Institut National des Science de l'Univers (INSU) of the Centre National de la Recherche Scientifique (CNRS) of France, and the University of Hawaii. 

% This research used the facilities of the Canadian Astronomy Data Centre operated by the National Research Council of Canada with the support of the Canadian Space Agency.

The Pan-STARRS1 Surveys (PS1) and the PS1 public science archive have been made possible through contributions by the Institute for Astronomy, the University of Hawaii, the Pan-STARRS Project Office, the MaxPlanck Society and its participating institutes, the Max Planck Institute for Astronomy, Heidelberg and the Max Planck Institute for Extraterrestrial Physics, Garching, The Johns Hopkins University, Durham University, the University of Edinburgh, the Queen’s University Belfast, the Harvard-Smithsonian Center for Astrophysics, the
Las Cumbres Observatory Global Telescope Network Incorporated, the National Central University of Taiwan, the Space Telescope Science Institute, the National Aeronautics and Space Administration under Grant No. NNX08AR22G issued through the Planetary Science Division of the NASA Science Mission Directorate, the National Science Foundation Grant No. AST-1238877, the University of Maryland, Eotvos Lorand University (ELTE), the Los Alamos National Laboratory, and the Gordon and Betty Moore Foundation. 

This work has made use of data from the European Space Agency (ESA) mission $\textit{Gaia}$ (\url{https://www.cosmos.esa.int/gaia}), processed by the $\textit{Gaia}$ Data Processing and Analysis Consortium (DPAC, \url{https://www.cosmos.esa.int/web/gaia/dpac/consortium}). Funding for the DPAC has been provided by national institutions, in particular the institutions participating in the Gaia Multilateral Agreement.

% The Acknowledgements section is not numbered. Here you can thank helpful
% colleagues, acknowledge funding agencies, telescopes and facilities used etc.

\section*{Data Availability}
The gold sample stars from the NGC~5466 stream are made available in the article and in its online supplementary material. A subset of the raw data underlying this article are publicly available via the Canadian Astronomical Data Center at \url{http://www.cadc-ccda.hia-iha.nrc-cnrc.gc.ca/en/megapipe/}. The remaining raw data and all processed data are available to members of the Canadian and French communities via reasonable requests to the principal investigators of the Canada-France Imaging Survey, Alan McConnachie and Jean-Charles Cuillandre. All data will be publicly available to the international community at the end of the proprietary period, scheduled for 2023.

\bibliographystyle{mnras}
\bibliography{MAIN} % if your bibtex file is called example.bib

\begin{thebibliography}{}
\makeatletter
\relax
\def\mn@urlcharsother{\let\do\@makeother \do\$\do\&\do\#\do\^\do\_\do\%\do\~}
\def\mn@doi{\begingroup\mn@urlcharsother \@ifnextchar [ {\mn@doi@}
  {\mn@doi@[]}}
\def\mn@doi@[#1]#2{\def\@tempa{#1}\ifx\@tempa\@empty \href
  {http://dx.doi.org/#2} {doi:#2}\else \href {http://dx.doi.org/#2} {#1}\fi
  \endgroup}
\def\mn@eprint#1#2{\mn@eprint@#1:#2::\@nil}
\def\mn@eprint@arXiv#1{\href {http://arxiv.org/abs/#1} {{\tt arXiv:#1}}}
\def\mn@eprint@dblp#1{\href {http://dblp.uni-trier.de/rec/bibtex/#1.xml}
  {dblp:#1}}
\def\mn@eprint@#1:#2:#3:#4\@nil{\def\@tempa {#1}\def\@tempb {#2}\def\@tempc
  {#3}\ifx \@tempc \@empty \let \@tempc \@tempb \let \@tempb \@tempa \fi \ifx
  \@tempb \@empty \def\@tempb {arXiv}\fi \@ifundefined
  {mn@eprint@\@tempb}{\@tempb:\@tempc}{\expandafter \expandafter \csname
  mn@eprint@\@tempb\endcsname \expandafter{\@tempc}}}

\bibitem[\protect\citeauthoryear{Ankerst, Breunig, peter Kriegel  \&
  Sander}{Ankerst et~al.}{1999}]{ankerst1999}
Ankerst M.,  Breunig M.~M.,  peter Kriegel H.,   Sander J.,  1999. ACM Press,
  pp 49--60

\bibitem[\protect\citeauthoryear{{Baumgardt}, {Hilker}, {Sollima}  \&
  {Bellini}}{{Baumgardt} et~al.}{2019}]{baumgardt2019}
{Baumgardt} H.,  {Hilker} M.,  {Sollima} A.,   {Bellini} A.,  2019, \mn@doi
  [\mnras] {10.1093/mnras/sty2997}, \href
  {https://ui.adsabs.harvard.edu/abs/2019MNRAS.482.5138B} {482, 5138}

\bibitem[\protect\citeauthoryear{{Bellazzini}, {Ibata}, {Malhan}, {Martin},
  {Famaey}  \& {Thomas}}{{Bellazzini} et~al.}{2020}]{bellazzini2020}
{Bellazzini} M.,  {Ibata} R.,  {Malhan} K.,  {Martin} N.,  {Famaey} B.,
  {Thomas} G.,  2020, \mn@doi [\aap] {10.1051/0004-6361/202037621}, \href
  {https://ui.adsabs.harvard.edu/abs/2020A&A...636A.107B} {636, A107}

\bibitem[\protect\citeauthoryear{{Belokurov}, {Evans}, {Irwin}, {Hewett}  \&
  {Wilkinson}}{{Belokurov} et~al.}{2006}]{belokurov2006}
{Belokurov} V.,  {Evans} N.~W.,  {Irwin} M.~J.,  {Hewett} P.~C.,   {Wilkinson}
  M.~I.,  2006, \mn@doi [\apjl] {10.1086/500362}, \href
  {https://ui.adsabs.harvard.edu/abs/2006ApJ...637L..29B} {637, L29}

\bibitem[\protect\citeauthoryear{{Bianchini}, {Ibata}  \& {Famaey}}{{Bianchini}
  et~al.}{2019}]{bianchini2019}
{Bianchini} P.,  {Ibata} R.,   {Famaey} B.,  2019, \mn@doi [\apjl]
  {10.3847/2041-8213/ab58d1}, \href
  {https://ui.adsabs.harvard.edu/abs/2019ApJ...887L..12B} {887, L12}

\bibitem[\protect\citeauthoryear{{Bonaca} \& {Hogg}}{{Bonaca} \&
  {Hogg}}{2018}]{bonaca2018}
{Bonaca} A.,  {Hogg} D.~W.,  2018, \mn@doi [\apj] {10.3847/1538-4357/aae4da},
  \href {https://ui.adsabs.harvard.edu/abs/2018ApJ...867..101B} {867, 101}

\bibitem[\protect\citeauthoryear{{Borsato}, {Martell}  \& {Simpson}}{{Borsato}
  et~al.}{2020}]{borsato2020}
{Borsato} N.~W.,  {Martell} S.~L.,   {Simpson} J.~D.,  2020, \mn@doi [\mnras]
  {10.1093/mnras/stz3479}, \href
  {https://ui.adsabs.harvard.edu/abs/2020MNRAS.492.1370B} {492, 1370}

\bibitem[\protect\citeauthoryear{{Boulade} et~al.,}{{Boulade}
  et~al.}{2003}]{boulade2003}
{Boulade} O.,  et~al., 2003, in {Iye} M.,  {Moorwood} A. F.~M.,  eds,  Society
  of Photo-Optical Instrumentation Engineers (SPIE) Conference Series Vol.
  4841, Instrument Design and Performance for Optical/Infrared Ground-based
  Telescopes. pp 72--81, \mn@doi{10.1117/12.459890}

\bibitem[\protect\citeauthoryear{{Bovy}}{{Bovy}}{2015}]{bovy2015}
{Bovy} J.,  2015, \mn@doi [\apjs] {10.1088/0067-0049/216/2/29}, \href
  {https://ui.adsabs.harvard.edu/abs/2015ApJS..216...29B} {216, 29}

\bibitem[\protect\citeauthoryear{{Boylan-Kolchin}, {Springel}, {White}  \&
  {Jenkins}}{{Boylan-Kolchin} et~al.}{2010}]{boylankolchin2010}
{Boylan-Kolchin} M.,  {Springel} V.,  {White} S. D.~M.,   {Jenkins} A.,  2010,
  \mn@doi [\mnras] {10.1111/j.1365-2966.2010.16774.x}, \href
  {https://ui.adsabs.harvard.edu/abs/2010MNRAS.406..896B} {406, 896}

\bibitem[\protect\citeauthoryear{{Carlin} \& {Sand}}{{Carlin} \&
  {Sand}}{2018}]{carlin2018}
{Carlin} J.~L.,  {Sand} D.~J.,  2018, \mn@doi [\apj]
  {10.3847/1538-4357/aad8c1}, \href
  {https://ui.adsabs.harvard.edu/abs/2018ApJ...865....7C} {865, 7}

\bibitem[\protect\citeauthoryear{{Carlin}, {Grillmair}, {Mu{\~n}oz}, {Nidever}
  \& {Majewski}}{{Carlin} et~al.}{2009}]{carlin2009}
{Carlin} J.~L.,  {Grillmair} C.~J.,  {Mu{\~n}oz} R.~R.,  {Nidever} D.~L.,
  {Majewski} S.~R.,  2009, \mn@doi [\apjl] {10.1088/0004-637X/702/1/L9}, \href
  {https://ui.adsabs.harvard.edu/abs/2009ApJ...702L...9C} {702, L9}

\bibitem[\protect\citeauthoryear{{Cautun} et~al.,}{{Cautun}
  et~al.}{2020}]{cautun2020}
{Cautun} M.,  et~al., 2020, \mn@doi [\mnras] {10.1093/mnras/staa1017}, \href
  {https://ui.adsabs.harvard.edu/abs/2020MNRAS.494.4291C} {494, 4291}

\bibitem[\protect\citeauthoryear{{Chambers} et~al.,}{{Chambers}
  et~al.}{2016}]{chambers2016}
{Chambers} K.~C.,  et~al., 2016, arXiv e-prints, \href
  {https://ui.adsabs.harvard.edu/abs/2016arXiv161205560C} {p. arXiv:1612.05560}

\bibitem[\protect\citeauthoryear{{Dalton} et~al.,}{{Dalton}
  et~al.}{2012}]{dalton2012}
{Dalton} G.,  et~al., 2012, in {McLean} I.~S.,  {Ramsay} S.~K.,   {Takami} H.,
  eds,  Society of Photo-Optical Instrumentation Engineers (SPIE) Conference
  Series Vol. 8446, Ground-based and Airborne Instrumentation for Astronomy IV.
  p. 84460P, \mn@doi{10.1117/12.925950}

\bibitem[\protect\citeauthoryear{{Deason}, {Belokurov}  \& {Evans}}{{Deason}
  et~al.}{2011}]{deason2011}
{Deason} A.~J.,  {Belokurov} V.,   {Evans} N.~W.,  2011, \mn@doi [\mnras]
  {10.1111/j.1365-2966.2011.19237.x}, \href
  {https://ui.adsabs.harvard.edu/abs/2011MNRAS.416.2903D} {416, 2903}

\bibitem[\protect\citeauthoryear{{Eilers}, {Hogg}, {Rix}  \& {Ness}}{{Eilers}
  et~al.}{2019}]{eilers2019}
{Eilers} A.-C.,  {Hogg} D.~W.,  {Rix} H.-W.,   {Ness} M.~K.,  2019, \mn@doi
  [\apj] {10.3847/1538-4357/aaf648}, \href
  {https://ui.adsabs.harvard.edu/abs/2019ApJ...871..120E} {871, 120}

\bibitem[\protect\citeauthoryear{Ester, Kriegel, Sander  \& Xu}{Ester
  et~al.}{1996}]{ester1996}
Ester M.,  Kriegel H.-P.,  Sander J.,   Xu X.,  1996. AAAI Press, pp 226--231

\bibitem[\protect\citeauthoryear{{Fardal}, {Huang}  \& {Weinberg}}{{Fardal}
  et~al.}{2015}]{fardal2015}
{Fardal} M.~A.,  {Huang} S.,   {Weinberg} M.~D.,  2015, \mn@doi [\mnras]
  {10.1093/mnras/stv1198}, \href
  {https://ui.adsabs.harvard.edu/abs/2015MNRAS.452..301F} {452, 301}

\bibitem[\protect\citeauthoryear{{Farrow} et~al.,}{{Farrow}
  et~al.}{2014}]{farrow2014}
{Farrow} D.~J.,  et~al., 2014, \mn@doi [\mnras] {10.1093/mnras/stt1933}, \href
  {https://ui.adsabs.harvard.edu/abs/2014MNRAS.437..748F} {437, 748}

\bibitem[\protect\citeauthoryear{{Fellhauer}, {Evans}, {Belokurov}, {Wilkinson}
   \& {Gilmore}}{{Fellhauer} et~al.}{2007}]{fellhauer2007}
{Fellhauer} M.,  {Evans} N.~W.,  {Belokurov} V.,  {Wilkinson} M.~I.,
  {Gilmore} G.,  2007, \mn@doi [\mnras] {10.1111/j.1365-2966.2007.12111.x},
  \href {https://ui.adsabs.harvard.edu/abs/2007MNRAS.380..749F} {380, 749}

\bibitem[\protect\citeauthoryear{{Gaia Collaboration} et~al.,}{{Gaia
  Collaboration} et~al.}{2016}]{gaia2016}
{Gaia Collaboration} et~al., 2016, \mn@doi [\aap]
  {10.1051/0004-6361/201629272}, \href
  {https://ui.adsabs.harvard.edu/abs/2016A&A...595A...1G} {595, A1}

\bibitem[\protect\citeauthoryear{{Gaia Collaboration} et~al.,}{{Gaia
  Collaboration} et~al.}{2018}]{gaia2018}
{Gaia Collaboration} et~al., 2018, \mn@doi [\aap]
  {10.1051/0004-6361/201833051}, \href
  {https://ui.adsabs.harvard.edu/abs/2018A&A...616A...1G} {616, A1}

\bibitem[\protect\citeauthoryear{{Gaia Collaboration}, {Brown}, {Vallenari},
  {Prusti}, {de Bruijne}, {Babusiaux}  \& {Biermann}}{{Gaia Collaboration}
  et~al.}{2020}]{gaia2020}
{Gaia Collaboration} {Brown} A.~G.~A.,  {Vallenari} A.,  {Prusti} T.,  {de
  Bruijne} J.~H.~J.,  {Babusiaux} C.,   {Biermann} M.,  2020, arXiv e-prints,
  \href {https://ui.adsabs.harvard.edu/abs/2020arXiv201201533G} {p.
  arXiv:2012.01533}

\bibitem[\protect\citeauthoryear{{Gravity Collaboration} et~al.,}{{Gravity
  Collaboration} et~al.}{2019}]{gravity2019}
{Gravity Collaboration} et~al., 2019, \mn@doi [\aap]
  {10.1051/0004-6361/201935656}, \href
  {https://ui.adsabs.harvard.edu/abs/2019A&A...625L..10G} {625, L10}

\bibitem[\protect\citeauthoryear{{Grillmair}}{{Grillmair}}{2009}]{grillmair2009}
{Grillmair} C.~J.,  2009, \mn@doi [\apj] {10.1088/0004-637X/693/2/1118}, \href
  {https://ui.adsabs.harvard.edu/abs/2009ApJ...693.1118G} {693, 1118}

\bibitem[\protect\citeauthoryear{{Grillmair} \& {Carlin}}{{Grillmair} \&
  {Carlin}}{2016}]{grillmair2016}
{Grillmair} C.~J.,  {Carlin} J.~L.,  2016, {Stellar Streams and Clouds in the
  Galactic Halo}.
p.~87, \mn@doi{10.1007/978-3-319-19336-6_4}

\bibitem[\protect\citeauthoryear{{Grillmair} \& {Johnson}}{{Grillmair} \&
  {Johnson}}{2006}]{grillmair2006}
{Grillmair} C.~J.,  {Johnson} R.,  2006, \mn@doi [\apjl] {10.1086/501439},
  \href {https://ui.adsabs.harvard.edu/abs/2006ApJ...639L..17G} {639, L17}

\bibitem[\protect\citeauthoryear{{Harris}}{{Harris}}{1996}]{harris1996}
{Harris} W.~E.,  1996, \mn@doi [\aj] {10.1086/118116}, \href
  {https://ui.adsabs.harvard.edu/abs/1996AJ....112.1487H} {112, 1487}

\bibitem[\protect\citeauthoryear{{Helmi}, {Veljanoski}, {Breddels}, {Tian}  \&
  {Sales}}{{Helmi} et~al.}{2017}]{helmi2017}
{Helmi} A.,  {Veljanoski} J.,  {Breddels} M.~A.,  {Tian} H.,   {Sales} L.~V.,
  2017, \mn@doi [\aap] {10.1051/0004-6361/201629990}, \href
  {https://ui.adsabs.harvard.edu/abs/2017A&A...598A..58H} {598, A58}

\bibitem[\protect\citeauthoryear{{Helmi}, {Babusiaux}, {Koppelman}, {Massari},
  {Veljanoski}  \& {Brown}}{{Helmi} et~al.}{2018}]{helmi2018}
{Helmi} A.,  {Babusiaux} C.,  {Koppelman} H.~H.,  {Massari} D.,  {Veljanoski}
  J.,   {Brown} A. G.~A.,  2018, \mn@doi [\nat] {10.1038/s41586-018-0625-x},
  \href {https://ui.adsabs.harvard.edu/abs/2018Natur.563...85H} {563, 85}

\bibitem[\protect\citeauthoryear{{Hernquist}}{{Hernquist}}{1990}]{hernquist1990}
{Hernquist} L.,  1990, \mn@doi [\apj] {10.1086/168845}, \href
  {https://ui.adsabs.harvard.edu/abs/1990ApJ...356..359H} {356, 359}

\bibitem[\protect\citeauthoryear{{Ibata}, {Lewis}, {Irwin}, {Totten}  \&
  {Quinn}}{{Ibata} et~al.}{2001}]{ibata2001}
{Ibata} R.,  {Lewis} G.~F.,  {Irwin} M.,  {Totten} E.,   {Quinn} T.,  2001,
  \mn@doi [\apj] {10.1086/320060}, \href
  {https://ui.adsabs.harvard.edu/abs/2001ApJ...551..294I} {551, 294}

\bibitem[\protect\citeauthoryear{{Ibata}, {Lewis}, {Irwin}  \&
  {Cambr{\'e}sy}}{{Ibata} et~al.}{2002}]{ibata2002}
{Ibata} R.~A.,  {Lewis} G.~F.,  {Irwin} M.~J.,   {Cambr{\'e}sy} L.,  2002,
  \mn@doi [\mnras] {10.1046/j.1365-8711.2002.05360.x}, \href
  {https://ui.adsabs.harvard.edu/abs/2002MNRAS.332..921I} {332, 921}

\bibitem[\protect\citeauthoryear{{Ibata} et~al.,}{{Ibata}
  et~al.}{2017a}]{ibata2017}
{Ibata} R.~A.,  et~al., 2017a, \mn@doi [\apj] {10.3847/1538-4357/aa855c}, \href
  {https://ui.adsabs.harvard.edu/abs/2017ApJ...848..128I} {848, 128}

\bibitem[\protect\citeauthoryear{{Ibata} et~al.,}{{Ibata}
  et~al.}{2017b}]{ibata_feh2017}
{Ibata} R.~A.,  et~al., 2017b, \mn@doi [\apj] {10.3847/1538-4357/aa8562}, \href
  {https://ui.adsabs.harvard.edu/abs/2017ApJ...848..129I} {848, 129}

\bibitem[\protect\citeauthoryear{{Ibata}, {Malhan}  \& {Martin}}{{Ibata}
  et~al.}{2019}]{ibata2019}
{Ibata} R.~A.,  {Malhan} K.,   {Martin} N.~F.,  2019, \mn@doi [\apj]
  {10.3847/1538-4357/ab0080}, \href
  {https://ui.adsabs.harvard.edu/abs/2019ApJ...872..152I} {872, 152}

\bibitem[\protect\citeauthoryear{{Ibata} et~al.,}{{Ibata}
  et~al.}{2020}]{ibata2020}
{Ibata} R.,  et~al., 2020, arXiv e-prints, \href
  {https://ui.adsabs.harvard.edu/abs/2020arXiv201205245I} {p. arXiv:2012.05245}

\bibitem[\protect\citeauthoryear{{Johnston}, {Hernquist}  \&
  {Bolte}}{{Johnston} et~al.}{1996}]{johnston1996}
{Johnston} K.~V.,  {Hernquist} L.,   {Bolte} M.,  1996, \mn@doi [\apj]
  {10.1086/177418}, \href
  {https://ui.adsabs.harvard.edu/abs/1996ApJ...465..278J} {465, 278}

\bibitem[\protect\citeauthoryear{{Johnston}, {Bullock}, {Sharma}, {Font},
  {Robertson}  \& {Leitner}}{{Johnston} et~al.}{2008}]{johnston2008}
{Johnston} K.~V.,  {Bullock} J.~S.,  {Sharma} S.,  {Font} A.,  {Robertson}
  B.~E.,   {Leitner} S.~N.,  2008, \mn@doi [\apj] {10.1086/592228}, \href
  {https://ui.adsabs.harvard.edu/abs/2008ApJ...689..936J} {689, 936}

\bibitem[\protect\citeauthoryear{{Juri{\'c}} et~al.,}{{Juri{\'c}}
  et~al.}{2008}]{juric2008}
{Juri{\'c}} M.,  et~al., 2008, \mn@doi [\apj] {10.1086/523619}, \href
  {https://ui.adsabs.harvard.edu/abs/2008ApJ...673..864J} {673, 864}

\bibitem[\protect\citeauthoryear{{King}}{{King}}{1962}]{king1962}
{King} I.,  1962, \mn@doi [\aj] {10.1086/108756}, \href
  {https://ui.adsabs.harvard.edu/abs/1962AJ.....67..471K} {67, 471}

\bibitem[\protect\citeauthoryear{{Kundu}, {Minniti}  \& {Singh}}{{Kundu}
  et~al.}{2019}]{kundu2019}
{Kundu} R.,  {Minniti} D.,   {Singh} H.~P.,  2019, \mn@doi [\mnras]
  {10.1093/mnras/sty3239}, \href
  {https://ui.adsabs.harvard.edu/abs/2019MNRAS.483.1737K} {483, 1737}

\bibitem[\protect\citeauthoryear{{K{\"u}pper}, {Balbinot}, {Bonaca},
  {Johnston}, {Hogg}, {Kroupa}  \& {Santiago}}{{K{\"u}pper}
  et~al.}{2015}]{kupper2015}
{K{\"u}pper} A. H.~W.,  {Balbinot} E.,  {Bonaca} A.,  {Johnston} K.~V.,  {Hogg}
  D.~W.,  {Kroupa} P.,   {Santiago} B.~X.,  2015, \mn@doi [\apj]
  {10.1088/0004-637X/803/2/80}, \href
  {https://ui.adsabs.harvard.edu/abs/2015ApJ...803...80K} {803, 80}

\bibitem[\protect\citeauthoryear{{Lamb}, {Venn}, {Shetrone}, {Sakari}  \&
  {Pritzl}}{{Lamb} et~al.}{2015}]{lamb2015}
{Lamb} M.~P.,  {Venn} K.~A.,  {Shetrone} M.~D.,  {Sakari} C.~M.,   {Pritzl}
  B.~J.,  2015, \mn@doi [\mnras] {10.1093/mnras/stu2674}, \href
  {https://ui.adsabs.harvard.edu/abs/2015MNRAS.448...42L} {448, 42}

\bibitem[\protect\citeauthoryear{{Laureijs} et~al.,}{{Laureijs}
  et~al.}{2011}]{laureijs2011}
{Laureijs} R.,  et~al., 2011, arXiv e-prints, \href
  {https://ui.adsabs.harvard.edu/abs/2011arXiv1110.3193L} {p. arXiv:1110.3193}

\bibitem[\protect\citeauthoryear{{Lehmann} \& {Scholz}}{{Lehmann} \&
  {Scholz}}{1997}]{lehmann1997}
{Lehmann} I.,  {Scholz} R.~D.,  1997, \aap, \href
  {https://ui.adsabs.harvard.edu/abs/1997A&A...320..776L} {320, 776}

\bibitem[\protect\citeauthoryear{{Leon}, {Meylan}  \& {Combes}}{{Leon}
  et~al.}{2000}]{leon2000}
{Leon} S.,  {Meylan} G.,   {Combes} F.,  2000, \aap, \href
  {https://ui.adsabs.harvard.edu/abs/2000A&A...359..907L} {359, 907}

\bibitem[\protect\citeauthoryear{{Lindegren} et~al.,}{{Lindegren}
  et~al.}{2018}]{lindegren2018}
{Lindegren} L.,  et~al., 2018, \mn@doi [\aap] {10.1051/0004-6361/201832727},
  \href {https://ui.adsabs.harvard.edu/abs/2018A&A...616A...2L} {616, A2}

\bibitem[\protect\citeauthoryear{{Luo} et~al.,}{{Luo} et~al.}{2015}]{luo2015}
{Luo} A.~L.,  et~al., 2015, \mn@doi [Research in Astronomy and Astrophysics]
  {10.1088/1674-4527/15/8/002}, \href
  {https://ui.adsabs.harvard.edu/abs/2015RAA....15.1095L} {15, 1095}

\bibitem[\protect\citeauthoryear{{Lux}, {Read}, {Lake}  \& {Johnston}}{{Lux}
  et~al.}{2012}]{lux2012}
{Lux} H.,  {Read} J.~I.,  {Lake} G.,   {Johnston} K.~V.,  2012, \mn@doi
  [\mnras] {10.1111/j.1745-3933.2012.01276.x}, \href
  {https://ui.adsabs.harvard.edu/abs/2012MNRAS.424L..16L} {424, L16}

\bibitem[\protect\citeauthoryear{{Mackereth} et~al.,}{{Mackereth}
  et~al.}{2019}]{mackereth2019}
{Mackereth} J.~T.,  et~al., 2019, \mn@doi [\mnras] {10.1093/mnras/sty2955},
  \href {https://ui.adsabs.harvard.edu/abs/2019MNRAS.482.3426M} {482, 3426}

\bibitem[\protect\citeauthoryear{{Majewski}, {Skrutskie}, {Weinberg}  \&
  {Ostheimer}}{{Majewski} et~al.}{2003}]{majewski2003}
{Majewski} S.~R.,  {Skrutskie} M.~F.,  {Weinberg} M.~D.,   {Ostheimer} J.~C.,
  2003, \mn@doi [\apj] {10.1086/379504}, \href
  {https://ui.adsabs.harvard.edu/abs/2003ApJ...599.1082M} {599, 1082}

\bibitem[\protect\citeauthoryear{{Malhan} \& {Ibata}}{{Malhan} \&
  {Ibata}}{2018}]{malhanibata2018}
{Malhan} K.,  {Ibata} R.~A.,  2018, \mn@doi [\mnras] {10.1093/mnras/sty912},
  \href {https://ui.adsabs.harvard.edu/abs/2018MNRAS.477.4063M} {477, 4063}

\bibitem[\protect\citeauthoryear{{Malhan} \& {Ibata}}{{Malhan} \&
  {Ibata}}{2019}]{malhanibata2019}
{Malhan} K.,  {Ibata} R.~A.,  2019, \mn@doi [\mnras] {10.1093/mnras/stz1035},
  \href {https://ui.adsabs.harvard.edu/abs/2019MNRAS.486.2995M} {486, 2995}

\bibitem[\protect\citeauthoryear{{Malhan}, {Ibata}  \& {Martin}}{{Malhan}
  et~al.}{2018}]{malhan2018}
{Malhan} K.,  {Ibata} R.~A.,   {Martin} N.~F.,  2018, \mn@doi [\mnras]
  {10.1093/mnras/sty2474}, \href
  {https://ui.adsabs.harvard.edu/abs/2018MNRAS.481.3442M} {481, 3442}

\bibitem[\protect\citeauthoryear{{Mateu}, {Read}  \& {Kawata}}{{Mateu}
  et~al.}{2018}]{mateu2018}
{Mateu} C.,  {Read} J.~I.,   {Kawata} D.,  2018, \mn@doi [\mnras]
  {10.1093/mnras/stx2937}, \href
  {https://ui.adsabs.harvard.edu/abs/2018MNRAS.474.4112M} {474, 4112}

\bibitem[\protect\citeauthoryear{{McConnachie} et~al.,}{{McConnachie}
  et~al.}{2018}]{mcconachie2018}
{McConnachie} A.~W.,  et~al., 2018, \mn@doi [\apj] {10.3847/1538-4357/aae8e7},
  \href {https://ui.adsabs.harvard.edu/abs/2018ApJ...868...55M} {868, 55}

\bibitem[\protect\citeauthoryear{{Miyamoto} \& {Nagai}}{{Miyamoto} \&
  {Nagai}}{1975}]{miyamoto1975}
{Miyamoto} M.,  {Nagai} R.,  1975, Publications of the Astronomical Society of
  Japan, \href {https://ui.adsabs.harvard.edu/abs/1975PASJ...27..533M} {27,
  533}

\bibitem[\protect\citeauthoryear{{Moreno}, {Pichardo}  \&
  {Vel{\'a}zquez}}{{Moreno} et~al.}{2014}]{moreno2014}
{Moreno} E.,  {Pichardo} B.,   {Vel{\'a}zquez} H.,  2014, \mn@doi [\apj]
  {10.1088/0004-637X/793/2/110}, \href
  {https://ui.adsabs.harvard.edu/abs/2014ApJ...793..110M} {793, 110}

\bibitem[\protect\citeauthoryear{{Navarro}, {Frenk}  \& {White}}{{Navarro}
  et~al.}{1996}]{navarro1996}
{Navarro} J.~F.,  {Frenk} C.~S.,   {White} S. D.~M.,  1996, \mn@doi [\apj]
  {10.1086/177173}, \href
  {https://ui.adsabs.harvard.edu/abs/1996ApJ...462..563N} {462, 563}

\bibitem[\protect\citeauthoryear{{Necib}, {Ostdiek}, {Lisanti}, {Cohen},
  {Freytsis}  \& {Garrison-Kimmel}}{{Necib} et~al.}{2019}]{necib2019}
{Necib} L.,  {Ostdiek} B.,  {Lisanti} M.,  {Cohen} T.,  {Freytsis} M.,
  {Garrison-Kimmel} S.,  2019, arXiv e-prints, \href
  {https://ui.adsabs.harvard.edu/abs/2019arXiv190707681N} {p. arXiv:1907.07681}

\bibitem[\protect\citeauthoryear{{Necib} et~al.,}{{Necib}
  et~al.}{2020}]{necib2020}
{Necib} L.,  et~al., 2020, \mn@doi [Nature Astronomy]
  {10.1038/s41550-020-1131-2}, \href
  {https://ui.adsabs.harvard.edu/abs/2020NatAs.tmp..137N} {}

\bibitem[\protect\citeauthoryear{{Odenkirchen} \& {Grebel}}{{Odenkirchen} \&
  {Grebel}}{2004}]{odenkirchen2004}
{Odenkirchen} M.,  {Grebel} E.~K.,  2004, in {Prada} F.,  {Martinez Delgado}
  D.,   {Mahoney} T.~J.,  eds,  Astronomical Society of the Pacific Conference
  Series Vol. 327, Satellites and Tidal Streams. p.~284 (\mn@eprint {arXiv}
  {astro-ph/0307481})

\bibitem[\protect\citeauthoryear{{Oliver}, {Elahi}, {Lewis}  \&
  {Power}}{{Oliver} et~al.}{2020}]{oliver2020}
{Oliver} W.~H.,  {Elahi} P.~J.,  {Lewis} G.~F.,   {Power} C.,  2020, arXiv
  e-prints, \href {https://ui.adsabs.harvard.edu/abs/2020arXiv201204823O} {p.
  arXiv:2012.04823}

\bibitem[\protect\citeauthoryear{{Pearson}, {K{\"u}pper}, {Johnston}  \&
  {Price-Whelan}}{{Pearson} et~al.}{2015}]{pearson2015}
{Pearson} S.,  {K{\"u}pper} A. H.~W.,  {Johnston} K.~V.,   {Price-Whelan}
  A.~M.,  2015, \mn@doi [\apj] {10.1088/0004-637X/799/1/28}, \href
  {https://ui.adsabs.harvard.edu/abs/2015ApJ...799...28P} {799, 28}

\bibitem[\protect\citeauthoryear{{Plummer}}{{Plummer}}{1911}]{plummer1911}
{Plummer} H.~C.,  1911, \mn@doi [\mnras] {10.1093/mnras/71.5.460}, \href
  {https://ui.adsabs.harvard.edu/abs/1911MNRAS..71..460P} {71, 460}

\bibitem[\protect\citeauthoryear{{Powell}}{{Powell}}{2013}]{powell2013}
{Powell} D.,  2013, \mn@doi [\nat] {10.1038/502022a}, \href
  {https://ui.adsabs.harvard.edu/abs/2013Natur.502...22P} {502, 22}

\bibitem[\protect\citeauthoryear{{Price-Whelan}}{{Price-Whelan}}{2017}]{pricewhelan2017}
{Price-Whelan} A.~M.,  2017, \mn@doi [The Journal of Open Source Software]
  {10.21105/joss.00388}, \href
  {https://ui.adsabs.harvard.edu/abs/2017JOSS....2..388P} {2, 388}

\bibitem[\protect\citeauthoryear{{Pryor}, {McClure}, {Fletcher}  \&
  {Hesser}}{{Pryor} et~al.}{1991}]{pryor1991}
{Pryor} C.,  {McClure} R.~D.,  {Fletcher} J.~M.,   {Hesser} J.~E.,  1991,
  \mn@doi [\aj] {10.1086/115929}, \href
  {https://ui.adsabs.harvard.edu/abs/1991AJ....102.1026P} {102, 1026}

\bibitem[\protect\citeauthoryear{{Racca} et~al.,}{{Racca}
  et~al.}{2016}]{racca2016}
{Racca} G.~D.,  et~al., 2016, in {MacEwen} H.~A.,  {Fazio} G.~G.,  {Lystrup}
  M.,  {Batalha} N.,  {Siegler} N.,   {Tong} E.~C.,  eds,  Society of
  Photo-Optical Instrumentation Engineers (SPIE) Conference Series Vol. 9904,
  Space Telescopes and Instrumentation 2016: Optical, Infrared, and Millimeter
  Wave. p. 99040O (\mn@eprint {arXiv} {1610.05508}),
  \mn@doi{10.1117/12.2230762}

\bibitem[\protect\citeauthoryear{{Sans Fuentes}, {De Ridder}  \&
  {Debosscher}}{{Sans Fuentes} et~al.}{2017}]{sansfuentes2017}
{Sans Fuentes} S.~A.,  {De Ridder} J.,   {Debosscher} J.,  2017, \mn@doi [\aap]
  {10.1051/0004-6361/201629719}, \href
  {https://ui.adsabs.harvard.edu/abs/2017A&A...599A.143S} {599, A143}

\bibitem[\protect\citeauthoryear{{Sch{\"o}nrich}, {Binney}  \&
  {Dehnen}}{{Sch{\"o}nrich} et~al.}{2010}]{schonrich2010}
{Sch{\"o}nrich} R.,  {Binney} J.,   {Dehnen} W.,  2010, \mn@doi [\mnras]
  {10.1111/j.1365-2966.2010.16253.x}, \href
  {https://ui.adsabs.harvard.edu/abs/2010MNRAS.403.1829S} {403, 1829}

\bibitem[\protect\citeauthoryear{{Searle} \& {Zinn}}{{Searle} \&
  {Zinn}}{1978}]{searle1978}
{Searle} L.,  {Zinn} R.,  1978, \mn@doi [\apj] {10.1086/156499}, \href
  {https://ui.adsabs.harvard.edu/abs/1978ApJ...225..357S} {225, 357}

\bibitem[\protect\citeauthoryear{{Sollima}}{{Sollima}}{2020}]{sollima2020}
{Sollima} A.,  2020, \mn@doi [\mnras] {10.1093/mnras/staa1209}, \href
  {https://ui.adsabs.harvard.edu/abs/2020MNRAS.495.2222S} {495, 2222}

\bibitem[\protect\citeauthoryear{{The Dark Energy Survey Collaboration}}{{The
  Dark Energy Survey Collaboration}}{2005}]{des2005}
{The Dark Energy Survey Collaboration} 2005, arXiv e-prints, \href
  {https://ui.adsabs.harvard.edu/abs/2005astro.ph.10346T} {pp
  astro--ph/0510346}

\bibitem[\protect\citeauthoryear{{Thomas}, {Famaey}, {Ibata}, {L{\"u}ghausen}
  \& {Kroupa}}{{Thomas} et~al.}{2017}]{thomas2017}
{Thomas} G.~F.,  {Famaey} B.,  {Ibata} R.,  {L{\"u}ghausen} F.,   {Kroupa} P.,
  2017, \mn@doi [\aap] {10.1051/0004-6361/201730531}, \href
  {https://ui.adsabs.harvard.edu/abs/2017A&A...603A..65T} {603, A65}

\bibitem[\protect\citeauthoryear{{Thomas} et~al.,}{{Thomas}
  et~al.}{2018a}]{thomas2018}
{Thomas} G.~F.,  et~al., 2018a, \mn@doi [\mnras] {10.1093/mnras/sty2604}, \href
  {https://ui.adsabs.harvard.edu/abs/2018MNRAS.481.5223T} {481, 5223}

\bibitem[\protect\citeauthoryear{{Thomas}, {Famaey}, {Ibata}, {Renaud},
  {Martin}  \& {Kroupa}}{{Thomas} et~al.}{2018b}]{thomas_stream2018}
{Thomas} G.~F.,  {Famaey} B.,  {Ibata} R.,  {Renaud} F.,  {Martin} N.~F.,
  {Kroupa} P.,  2018b, \mn@doi [\aap] {10.1051/0004-6361/201731609}, \href
  {https://ui.adsabs.harvard.edu/abs/2018A&A...609A..44T} {609, A44}

\bibitem[\protect\citeauthoryear{{Thomas} et~al.,}{{Thomas}
  et~al.}{2019}]{thomas2019}
{Thomas} G.~F.,  et~al., 2019, \mn@doi [\apj] {10.3847/1538-4357/ab4a77}, \href
  {https://ui.adsabs.harvard.edu/abs/2019ApJ...886...10T} {886, 10}

\bibitem[\protect\citeauthoryear{{Thomas} et~al.,}{{Thomas}
  et~al.}{2020}]{thomas2020}
{Thomas} G.~F.,  et~al., 2020, arXiv e-prints, \href
  {https://ui.adsabs.harvard.edu/abs/2020arXiv200904487T} {p. arXiv:2009.04487}

\bibitem[\protect\citeauthoryear{{Venn}, {Tolstoy}, {Kaufer}  \&
  {Kudritzki}}{{Venn} et~al.}{2004}]{venn2004}
{Venn} K.~A.,  {Tolstoy} E.,  {Kaufer} A.,   {Kudritzki} R.~P.,  2004, in
  {McWilliam} A.,  {Rauch} M.,  eds, Origin and Evolution of the Elements.
  p.~58 (\mn@eprint {arXiv} {astro-ph/0305188})

\bibitem[\protect\citeauthoryear{{White} \& {Rees}}{{White} \&
  {Rees}}{1978}]{white1978}
{White} S.~D.~M.,  {Rees} M.~J.,  1978, \mn@doi [\mnras]
  {10.1093/mnras/183.3.341}, \href
  {https://ui.adsabs.harvard.edu/abs/1978MNRAS.183..341W} {183, 341}

\bibitem[\protect\citeauthoryear{{York} et~al.,}{{York}
  et~al.}{2000}]{york2000}
{York} D.~G.,  et~al., 2000, \mn@doi [\aj] {10.1086/301513}, \href
  {https://ui.adsabs.harvard.edu/abs/2000AJ....120.1579Y} {120, 1579}

\makeatother
\end{thebibliography}

% \begin{thebibliography}{99}
% \bibitem[\protect\citeauthoryear{Author}{2012}]{Author2012}
% Author A.~N., 2013, Journal of Improbable Astronomy, 1, 1
% \bibitem[\protect\citeauthoryear{Others}{2013}]{Others2013}
% Others S., 2012, Journal of Interesting Stuff, 17, 198
% \end{thebibliography}

%%%%%%%%%%%%%%%%%%%%%%%%%%%%%%%%%%%%%%%%%%%%%%%%%%

%%%%%%%%%%%%%%%%% APPENDICES %%%%%%%%%%%%%%%%%%%%%

\appendix

\bsp	% typesetting comment
\label{lastpage}
\end{document}